\documentclass[10pt,twocolumn,amsmath,amssymb,osajnl,floatfix,groupedaddress]{revtex4-1}

\usepackage[utf8]{inputenc}

\usepackage{graphicx}
\usepackage{dcolumn}
\usepackage{bm}
\usepackage{color}
\usepackage{txfonts}
\usepackage{microtype}
\usepackage{overpic}
\usepackage[normalem]{ulem}

\usepackage{siunitx}
\DeclareSIUnit{\atomicunit}{a.u.}

\begin{document}

\author{B. Willenberg}
\author{J. Maurer}\email{jocmaure@phys.ethz.ch}
\author{U. Keller}
\affiliation{Department of Physics, ETH Zurich, 8093, Zurich, Switzerland}

\author{J. Dan\v{e}k}
\author{M. Klaiber}
\author{N. Teeny}
\author{K. Z. Hatsagortsyan}\email{k.hatsagortsyan@mpi-k.de}
\author{C. H. Keitel}
\affiliation{Max-Planck-Institut f\"ur Kernphysik, Saupfercheckweg 1, 69117 Heidelberg, Germany}

\bibliographystyle{apsrev4-1}

\title{Holographic interferences in strong-field ionization beyond the dipole approximation:\\
The influence of the peak and focal volume averaged laser intensity}

\date{\today}

\begin{abstract}

In strong-field ionization interferences between electron trajectories create a variety of interference structures in the final momentum distributions. Among them, the interferences between electron pathways that are driven directly to the detector and the ones that rescatter significantly with the parent ion lead to holography-type interference patterns that received great attention in recent years. 
In this work, we study the influence of the magnetic field component onto the holographic interference pattern, an effect beyond the electric dipole approximation, in experiment and theory. The experimentally observed nondipole signatures are analyzed via quantum trajectory Monte Carlo simulations. 
We provide explanations for the experimentally demonstrated asymmetry in the holographic interference pattern and its non-uniform photoelectron energy dependence as well as for the variation of the topology of the holography-type interference pattern along the laser field direction.
Analytical scaling laws of the interference features are derived, and their direct relation to either the focal volume averaged laser intensities, or to the peak intensities are identified. The latter, in particular, provides a direct access to the peak intensity in the focal volume.
\end{abstract}

\maketitle

\section{Introduction}
Recently, holographic interferences were observed in strong-field ionization of atoms and molecules. They have the potential to provide information about the target and the ionization process with attosecond time- and \aa{}ngstr{\"o}m spatial-resolution \cite{Huismans_2011,Bian_2011,Marchenko_2011,Huismans_2012}. 
The original concept of holography is based on the interference of two waves: a direct reference wave and a signal wave that scattered off the target. The information about the target is encoded in the interference pattern of the two waves.
The holographic interference pattern from strong-field ionization is contained in the photoelectron momentum distribution (PMD) and is based on the recollision concept \cite{Corkum_1993}. The reference beam of the holography scheme \cite{Stroke_1966} is the directly ionized electron wave packet, while the signal beam is the electron wave packet scattered off the ionic core during recollision. This concept of strong-field holography enables to extract time-resolved information on the underlying electron dynamics \cite{Hickstein_2012,Zhou_2016,He_2018} and on the molecular structure \cite{Bian_2012,Bian_2014,Meckel_2014,Haertelt_2016,Walt_2017}.

Strong-field holography has been commonly applied in the regime of the electric dipole approximation. However, aiming at increased resolution of the holographic interference pattern, shorter de-Broglie wavelengths of the sampling electron wave, i.e. larger recollision energies of the electron, are required. This can be achieved by an increase of the ponderomotive energy $U_p$ of the electron in the laser field. Thereby nondipole effects related to the magnetic field of the laser become important for the description of holographic measurements \cite{Chelkowski_2015,Ivanov_2016,Brennecke_2018,Brennecke_2018a,Brennecke_2019}.

The leading nondipole effect for the continuum electron in the strong-field ionization process is a drift along the laser propagation direction. This forward drift has been measured cycle averaged \cite{Smeenk_2011}, sub-cycle time resolved \cite{Willenberg_2019} and theoretically analyzed in Ref.\cite{Titi_2012,Klaiber_2013c,Chelkowski_2015,Cricchio_2015,Chelkowski_2017,He_2017,Chelkowski_2018}. The drift induced by the laser magnetic field is known to reduce the probability of  electron recollision with the parent ion \cite{Dammasch_2001,Walser_2000a,Milosevic_2000,Chirila_2002,Klaiber_2005,Kohler_2012b}.

As demonstrated in a recent experiment \cite{Ludwig_2014} the drift induced by the laser magnetic field affects recollisions with the parent ion and modifies the Coulomb focusing, resulting from the electron multiple forward scatterings \cite{Brabec_1996}. The breakdown of the dipole approximation has been manifested in the counterintuitive shift of the peak of the photoelectron distribution against the laser propagation direction, which is due to the interplay between the nondipole and Coulomb field effects \cite{Foerre_2006,Keil_2017,Tao_2017,Danek_2018},  observed also in elliptically polarized light \cite{Maurer_2018,Danek_2018b}. The shift of the photoelectron distribution ridge along the laser propagation direction is negative for low energy electrons and positive for high energy ones \cite{Danek_2018}. Similar asymmetries have been predicted theoretically for the strong-field holography pattern \cite{Chelkowski_2015,Ivanov_2016,Brennecke_2018,Brennecke_2018a}.

In this paper we report the first experimental observation of nondipole signatures in the photoelectron holographic interference pattern from strong-field ionization. The onset of relativistic (nondipole) effects is expected at high laser intensities and long wavelengths \cite{Palaniyappan_2006a,Reiss_2008,Klaiber_2017}, which have been investigated  in experiments with highly charged ions \cite{McNaught_1997,Moore_1999,Dammasch_2001,Chowdhury2003, Gubbini_2005,Palaniyappan2005,DiChiara_2008,Palaniyappan_2008,Ekanayake_2013}.
However, the precision of the presented measurement allows us to observe nondipole features with nonrelativistic laser intensities, when the nondipole momentum shift in the laser propagation direction is smaller by an order of magnitude with respect to the width of the transverse PMD.
The holography pattern in the PMD shows an asymmetry with respect to the laser propagation direction in this regime. Our theoretical description using quantum trajectory Monte Carlo (QTMC) simulations confirms and explains the observed asymmetry. 
The analytical scaling laws of the observed features are derived. We show how the interplay between the momentum transfer due to the laser radiation pressure and Coulomb focusing lead to a nonuniform distortion of the holography pattern.

\begin{figure}
	\centering
	\includegraphics[width=1.0\columnwidth]{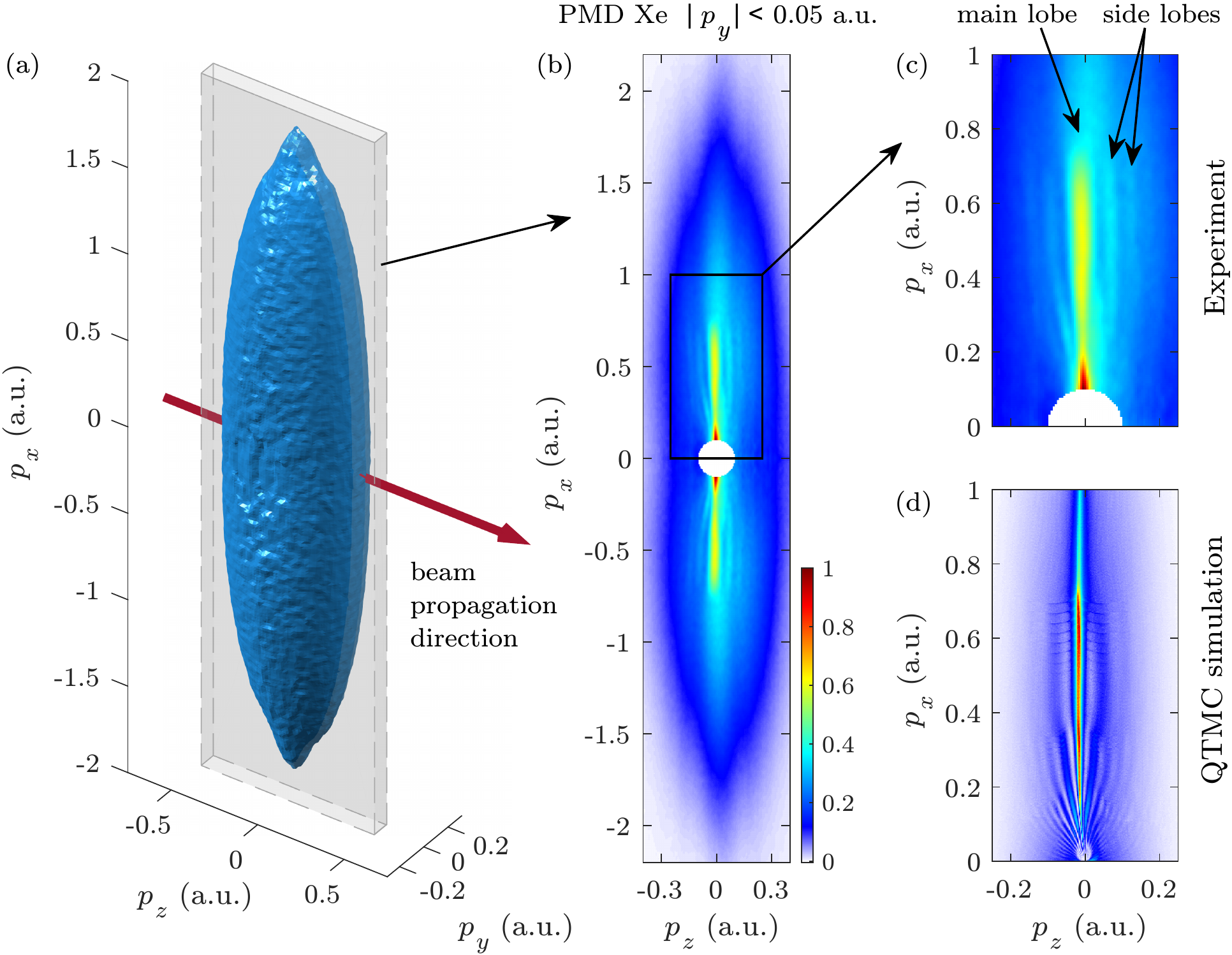}
	\caption{Experimental PMD from strong-field ionization of xenon atoms: (a) Tomographic reconstruction of the 3D PMD, (b) measured PMD slice within $|p_y|\leq\SI{0.05}{\atomicunit}$, (c) PMD enlarged in, with removal of the zero energy spot due to electrons captured in Rydberg states. A 6-cycle  laser pulse of linear polarization is used with an intensity of \SI{5.8e13}{\watt\per\square\centi\meter}, and  a wavelength of \SI{3.4}{\micro\meter}, (d) corresponding semiclassical simulation.}
	\label{fig:measurement}
\end{figure}

The structure of the presented work is the following. In Sec.~\ref{exp} the experimental details and observations are discussed. Our theoretical model is introduced in Sec.~\ref{theo}. The analysis of the obtained results is carried out in Sec.~\ref{analysis}, followed by a  conclusion in Sec.~\ref{conclusion}. Atomic units are used throughout.

\section{Experiment}\label{exp}
We access the weakly relativistic nondipole regime of strong-field ionization with linearly polarized mid-infrared pulses generated by an OPCPA-system. The system  delivers \SI{45}{\femto\second} long pulses at a central wavelength of \SI{3.4}{\micro\meter} with a pulse energy of up to \SI{21.8}{\micro\joule} at a repetition rate of \SI{50}{\kilo\hertz} \cite{Mayer_2014OSA,Mayer_2013}. The pulses are tightly focused in a back-focusing geometry by a dielectric mirror into the gas target. 
The PMDs from strong-field ionization with few-cycle-mid-IR pulses were recorded with a velocity map imaging spectrometer (VMIS) \cite{Eppink_1997,Parker_1997}. 
We ensured that the dielectric mirror does not induce any significant asymmetries along the beam propagation direction with reference PMDs recorded at a wavelength of 800\,nm.     

The VMIS can only image projections of the full 3D PMD onto the detector plane. In those projections the interference features are partially washed out and the kinetic energy spectra of the photoelectrons cannot be accessed directly.
In order to enhance the visibility of the interference patterns we consider cuts through the full 3D PMD.
We obtain the full 3D PMD from tomographic reconstruction \cite{Wollenhaupt_2009, Smeenk_2009momentum, Dimitrovski_2014} as the commonly used Abel inversion cannot be applied in our case: The PMD does not feature the needed cylindrical symmetry in the case of strong-field ionization beyond the limit of the electric dipole approximation.
We recorded PMDs at a peak intensity of $~ \SI{5e13}{\watt\per\square\centi\meter}$ in angle steps of \ang{1} covering the full range $[\ang{0},\ang{180})$ that is required to obtain the reconstructed 3D PMD with the required resolution. The full 3D PMD was reconstructed with the filtered back-projection algorithm. Subsequently we choose the region of $|p_y|<\SI{0.05}{\atomicunit}$ and project it onto the $p_x$-$p_z$-plane (Fig. \ref{fig:measurement}).

The momentum zero is calibrated according to Ref. \cite{Ludwig_2014}, namely by projecting a thin slice of $|p_x|<\SI{0.01}{\atomicunit}$ onto the beam propagation axis and fitting this distribution as a function of $p_z$ with a Lorentzian profile. The peak of this distribution is dominated by electrons stemming from atoms that are left in a highly excited, but neutral state after the laser pulse and that were subsequently ionized by the DC field of the spectrometer and end up centered around zero momentum \cite{Smeenk_2011, Nubbemeyer_2008, Eichmann_2009}.

Although our experimental parameters are in a regime that is considered the weakly relativistic regime we observe a significant influence of the laser magnetic field in the holography pattern of the photoelectron. 
The definition of the coordinate system is illustrated in Fig.~\ref{fig:measurement}: The laser beam propagates in positive $z$-direction and the laser field is polarized linearly along the $x$-direction. The momentum space coordinates ($p_i$) are co-aligned with the corresponding spatial coordinates.
Throughout this article we focus on cuts, i.e. projections onto the $p_x$-$p_z$-plane from a range of $|p_y|<\SI{0.05}{\atomicunit}$ (Fig.~\ref{fig:measurement}).

\begin{figure*}
	\centering
	\includegraphics[width=0.9\textwidth]{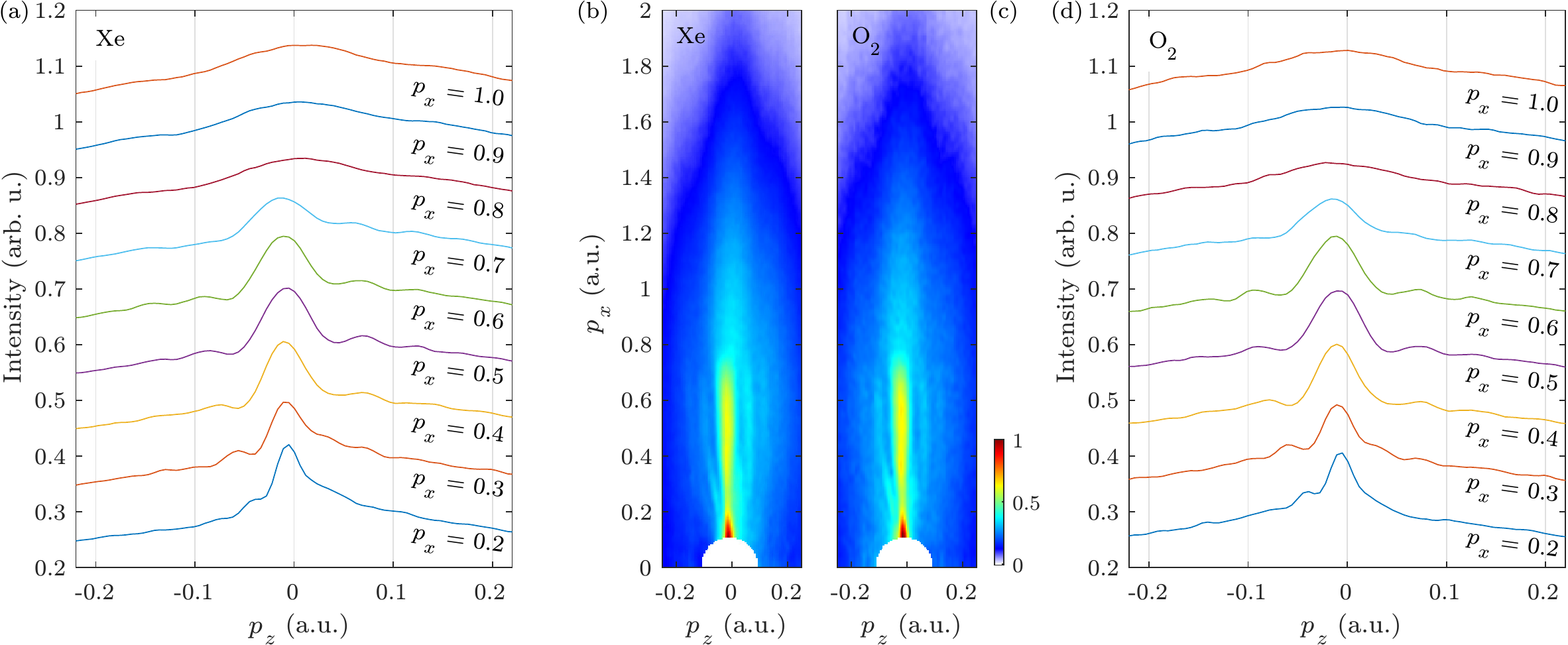}
	\caption{Lineouts of the experimental PMDs from strong-field ionization at linear polarization of the two targets xenon and oxygen at different $p_x$ positions along the laser polarization axis (in atomic units). The integration band for the lineouts in $p_x$ direction has a width of \SI{0.04}{\atomicunit}. The linouts are normalized to the integrated signal and for clarity of the figure offset in absolut intensity by the value of the $p_x$ position. Both targets show a very similar holography pattern with cutoffs for the number of interference maxima around $p_x = \SI{0.35}{\atomicunit}$ and $p_x = \SI{0.75}{\atomicunit}$ and an asymmetry in the intensity of the corresponding left and right sidelobe peak in laser beam propagation direction $p_z$ (for details see text).}
	\label{fig:O2vsXe}
\end{figure*}

The main target studied in this work is xenon with an ionization potential of $I_p=\SI{12.13}{\electronvolt}$. We performed additional measurements with a diatomic molecule, oxygen ($I_p=\SI{12.07}{\electronvolt}$), with an ionization potential close to the one of xenon to support the general nature of our findings from xenon (Fig.~\ref{fig:O2vsXe}).

In the experimental momentum images we observe a main lobe around $p_z \approx \SI{0}{\atomicunit}$, accompanied by additional lobes on both sides. Both, the main lobe and the sidelobes show an asymmetry along $p_z$. The asymmetry is especially visible in the lineouts created from the cut (Fig.~\ref{fig:O2vsXe}). For the lineouts, we projected slices centred at fixed $p_x$ of a width of \SI{0.04}{\atomicunit} in $p_x$ on the $p_z$ axis. The lineouts show, that the main ridge as well as the sidelobes are offset towards negative $p_z$-momenta. The right sidelobes are stronger in intensity than the left ones.

The difference of the data from the two targets is marginal, showing that the observations are mostly independent of the initial state.
Accordingly, the subsequent analysis and theoretical description focuses on the propagation of the photoelectron in the continuum.

\section{Trajectory-based semiclassical model}
\label{theo}
The observed holographic interferences can be qualitatively described by the perturbative nondipole strong-field approximation (SFA) \cite{Klaiber_2005}, where the rescattering is treated as a perturbation. 
However, a quantitative correct picture, including the effect of Coulomb focusing, can be obtained only with a nonperturbative treatment of the Coulomb potential of the ionic core. 
Our theoretical treatment is based on 3D QTMC simulations. In the latter the ionized electron wave packet is formed at the tunnel exit according to tunnel ionization theory \cite{PPT,ADK}, and further propagated in the continuum via the classical equations of motion in the laser and Coulomb field of the atomic core. In the QTMC simulation a phase is attached to each trajectory, which accrues along the trajectory and allows to describe quantum interferences between trajectories \cite{Li_2014a,Shvetsov-Shilovski_2016}. The photoelectron momentum distribution is calculated as the coherent sum over all trajectories
\begin{eqnarray}
	|M(\mathbf{p})|^2=\lim_{ t\rightarrow \infty}\left|\sum_l \sqrt{W\left(\mathbf{p}_i^{(l)}\right)}\exp[i  {\cal S}^{(l)}(t_i^{(l)},\mathbf{p})] \right|^2,
	\label{sum_s}
\end{eqnarray}
where $W(\mathbf{p}_i^{(l)})$ is the tunnel ionization probability of the electron with the initial  momentum $\mathbf{p}_i^{(l)}$ at the tunnel exit.  The summation in Eq.~(\ref{sum_s}) is carried out over all possible trajectories [labelled by $(l)$] in the laser and Coulomb field that  start at the ionization time $t_i^{(l)}$ at the tunnel exit $\mathbf{r}_i^{(l)}$,
with the initial momentum $\mathbf{p}_i^{(l)}=(0,p_{yi}^{(l)},p_{zi}^{(l)})$, and end up asymptotically at a given final momentum $\mathbf{p}$ of the photoelectron. The tunnel exit  is derived using the quasistatic model of Ref. \cite{Pfeiffer_2012}, including the Coulomb field of the atomic core, and the static atomic polarizability. The trajectories are found numerically solving Newton equations in the relativistic formulation:
\begin{eqnarray}
\frac{d \textbf{p}}{dt}=-\textbf{E}(\phi)-\textbf{v}\times \textbf{B}(\phi)-\nabla V(\textbf{r}),
\end{eqnarray}
with the laser electric $\textbf{E}(\phi)$, and magnetic $\textbf{B}(\phi)$ fields, respectively, the laser phase $\phi=\omega(t-z/c)$, and the potential of the atomic core $V(\textbf{r})$. 

The phase of the $l^{th}$-trajectory $ {\cal S}^{(l)}(t,\mathbf{p})$ is determined by the classical action along the trajectory $\textbf{r}^{(l)}(t)$ in the laser and Coulomb field \cite{Popov_2005}:
\begin{eqnarray}
{\cal S}^{(l)}(t,\mathbf{p})&=&-\mathbf{p}\cdot\mathbf{r}^{(l)}(\infty)+\int^{\infty}_{t}d\tau\,\left[{\cal L}^{(l)}(\tau)-(c^2-I_p)\right]  ,\nonumber\\
\label{Trajectory_phase}
\end{eqnarray}
where $\textbf{p}$ is the final photoelectron momentum,  and ${\cal L}^{(l)}$ is the relativistic Lagrangian of the electron in the laser and Coulomb field \cite{Landau_2}:  
\begin{eqnarray}
{\cal L}^{(l)}=-c^2\sqrt{1- \textbf{v}^2(t)/c^2}-\textbf{A}(\phi)\cdot\textbf{v}(t)+\varphi (\phi) +V(\mathbf{r}(t)),
\end{eqnarray}
with  the time-dependent electron coordinate $\textbf{r}(t)$, and velocity  $\textbf{v}(t)$  along the trajectory. The laser field propagating along the $z$-axis, with the electric and magnetic field along the $x$- and $y$-axis, respectively, is described by the  potentials in the Göppert-Mayer gauge: $\textbf{A}(\phi)=-\hat{\mathbf{z}} (x E(\phi))/c$ and $\varphi=-xE(\phi)$ \cite{Reiss_1992}. Using the electron equation of motion, the classical action can be represented as
\begin{eqnarray}
{\cal S}^{(l)}(t,\mathbf{p})&=& -\textbf{p}_i \cdot \textbf{r}_i - \int^{\infty}_{t}d\tau\left\{ \varepsilon(\tau) +I_p +V(\textbf{r}(\tau))\right.\nonumber \\
&&\left. - \textbf{r}(\tau)\cdot \nabla V(\textbf{r}(\tau))-\frac{z(\tau)}{c}\textbf{v}(\tau)\cdot \textbf{E}(\tau) \right\},
\label{phase}
\end{eqnarray}
with the kinetic energy $\varepsilon =c^2(\gamma  -1)$, the Lorentz  $\gamma$-factor, the initial coordinate $\textbf{r}_i$ and momentum $\textbf{p}_i$. Note that $\textbf{p}_i \cdot \textbf{r}_i=0$ in the  tunneling regime as the electron longitudinal momentum along the field is vanishing at the tunnel exit. 

The full QTMC simulation for the given parameters is presented in Fig.~\ref{fig:ctmctra}. It incorporates all possible trajectories, as well as  focal volume averaging. 
To elucidate the contribution of the different type of trajectories in forming the holography structure, we also carried out a model simulation where we included the two main type trajectories, only (Fig.~\ref{fig:ctmctra}). When all trajectories are included, the momentum distribution becomes more smooth, however, the main features of the interference structure are already given by the spectra based on the two main trajectories. The PMD reveals interference structures with a middle lobe and with side wings, that qualitatively were already known from the nonrelativistic regime \cite{Huismans_2011}. The nondipole effects shift the holography interference pattern to the negative $p_z$ direction breaking the symmetry with respect to $p_z$. The Coulomb field from the parent ion squeezes the interference structure in the whole transverse plane.

The focal averaging is carried out by incoherent superposition of PMDs from QTMC simulations for 10 different intensities. Each intensity is weighted by the factor $F(I)=-dV/dI$ according to the weight of the partial focal volume \cite{Augst_1991,Brichta_2006}, calculated for the paraxial model of a focused laser beam: 
$V= [2\pi^2w_0^4/(3\lambda)]\left(2C+C^3/3 -2\tan^{-1}C\right)$,
with $C=\sqrt{I_0/I-1}$, the instantaneous intensity $I$, and the peak intensity $I_0$ . 

We would like to point out that we analyzed in the QTMC simulations two types of the effective potentials for the atomic core (xenon singly charged ion) \cite{Rogers_1981,Milosevic_2009} and the polarizability of the atom. However, we could not find any significant influence on the holographic pattern as the recollisions take place fairly far from the atomic core.
  
\section{Analysis and discussion of the interference pattern} \label{analysis}
\subsection{Classification of trajectories}

\begin{figure}
	\centering
	\includegraphics[width=\columnwidth]{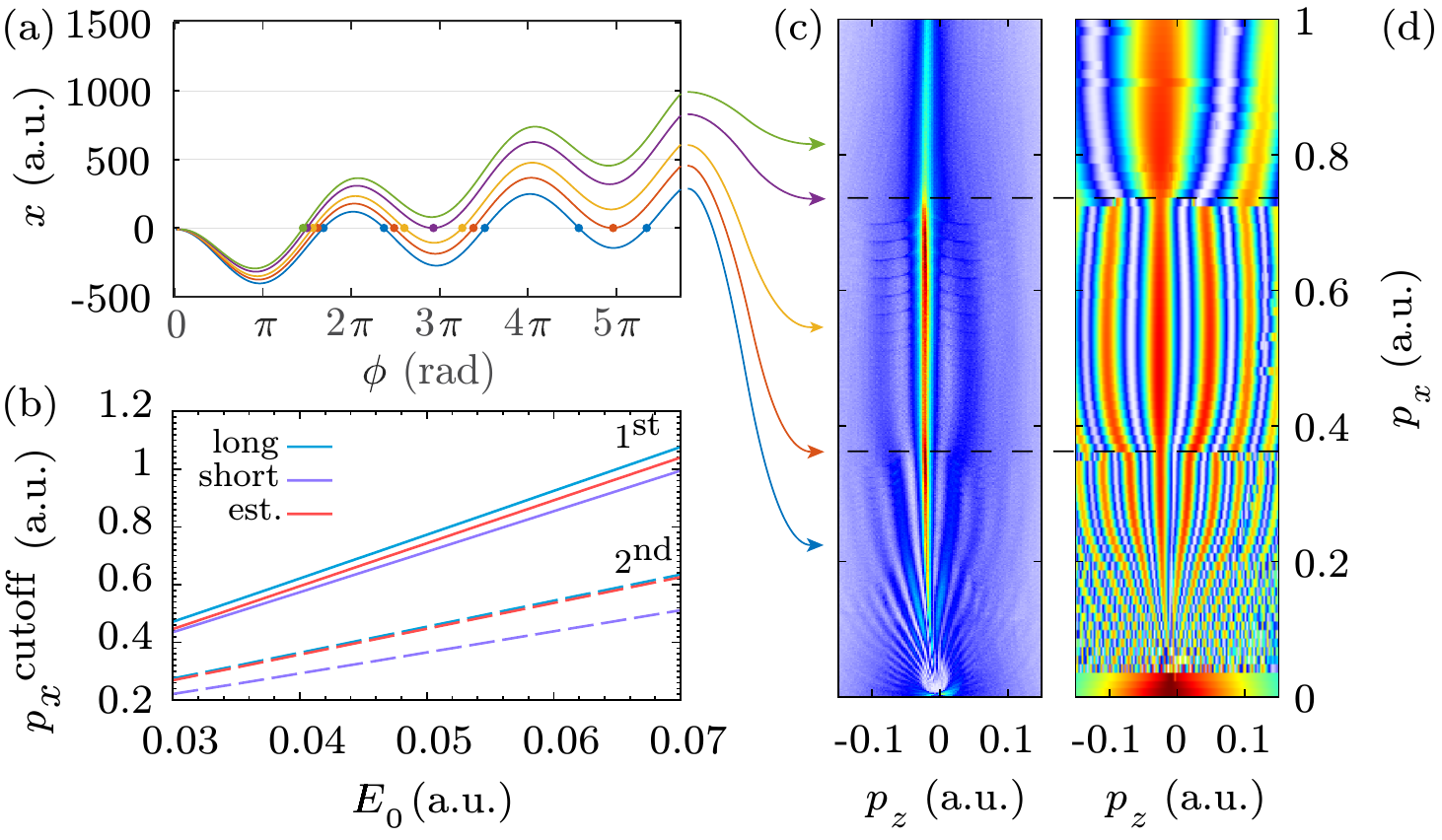}
	\caption{(a) Schematic examples of recolliding trajectories:  the trajectory with a single recollision (recollisions are indicated by dots) exists for $0.2E_0/\omega<|p_x|<E_0/\omega$ (green); the second and third returns exists for trajectories up to  $0.1E_0/\omega<|p_x|< 0.2E_0/\omega$  (yellow),  four or more recollisions exist at $|p_x|<0.1E_0/\omega$ (blue), and so on, see Eqs.~(\ref{pxcutoff1_})-(\ref{pxcutoff2_}). The momentum transfer at a recollision is enhanced for the specific trajectories (purple and red) when  the recollision velocity is vanishing (slow recollision) \cite{Liu_2011,Kastner_2012}. (b) The smallest $p_x$ cutoff for trajectories with a single recollision (solid), corresponding to the 1st slow recollision, and three recollisions (dashed), corresponding to the 2nd slow recollision; (long) The  slow recollision condition in a monochromatic laser field; (short) The slow recollision conditions in a 50 fs laser pulse,  and (est.) The analytical estimates of Eqs.~(\ref{pxcutoff1_}) and (\ref{pxcutoff2_}). The cutoffs are clearly visible in the PMDs from the QTMC (c) and simplified model simulations (d). For the simulations the laser and atom parameters are the same as in Fig.~\ref{fig:measurement}.
	}
	\label{fig:ctmctra}
\end{figure}
\begin{figure}
	\centering
	\includegraphics[width=0.85\columnwidth,trim=0.15cm 0cm 0cm 0.25cm,clip]{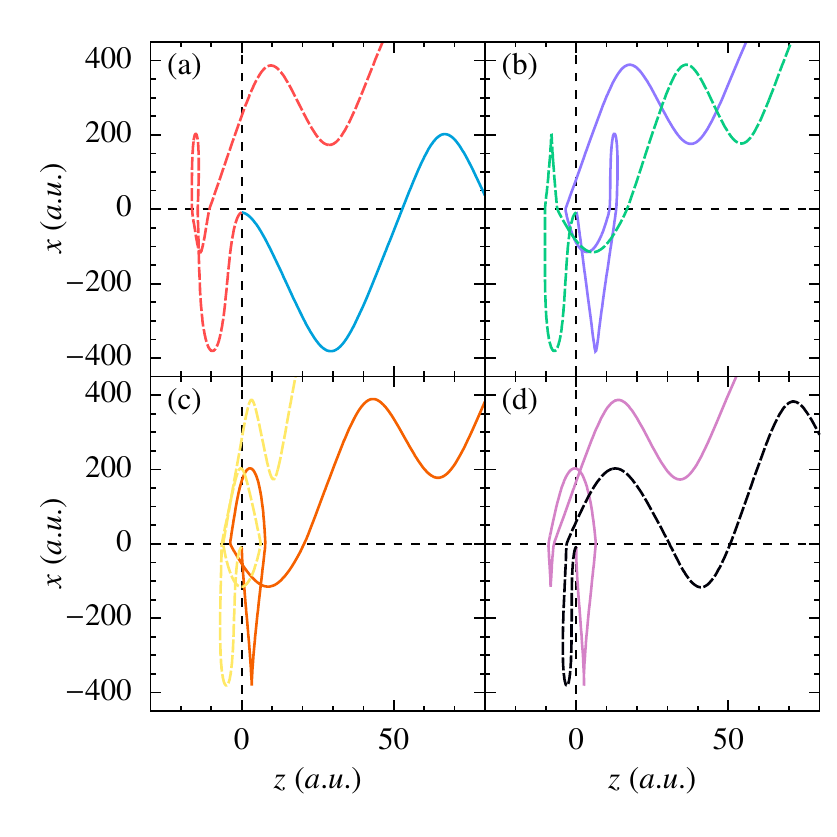}\\
   	\includegraphics[width=0.85\columnwidth,trim=0.15cm 0.2cm 0cm 0cm,clip]{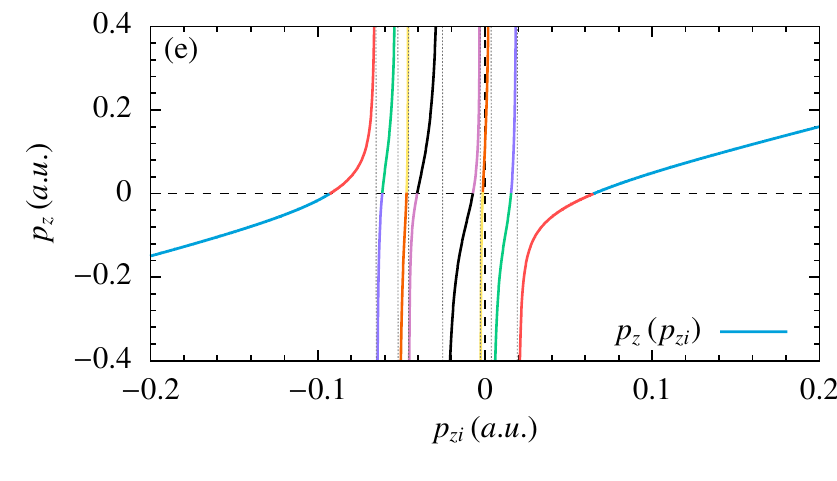}
	\caption{The examples of 8 types of trajectories identified in the QTMC simulation at $p_x=0.5$: (a) Trajectories with no significant rescattering (blue), and with a significant rescattering at the  3rd recollision (red), (b) Trajectories with a significant rescattering at the  2nd recollision (green), and  at the 2nd or 3rd recollisions (violet), (c)  at all recollisions (yellow), or at the 1st and 2nd recollisions (coral), (d) at the 1st (black), or at  1st and 3rd recollisions (magenta). (e) The final transverse momentum, $p_{z}$ vs the initial, $p_{zi}$, which determined the type of trajectories indicated by color. The color of trajectories in panels (a)-(d) indicate the type according to the panel (e). The type of pair trajectories in each of panels (a)-(d) is interchanged with continuous variation of $p_{zi}$ (e.g. blue-type of trajectory becomes red-type at decreasing $p_{zi}$, at $p_{zi}>0$).  }
	\label{tra2}
\end{figure}

The interference pattern in the strong-field holography is due to interference of several paths with close ionization times, along which the electron scatters forward at recollisions, and finally yields the same asymptotic momentum \cite{Huismans_2012}. It is known that the inter-cycle  interference induce a horizontal interference structure (perpendicular to $p_x$) \cite{Arbo_2006a}, which is usually not observed in experiments because of focal volume averaging and is neglected in our consideration. The interference (inter-half-cycle) of short and long trajectories  also induces horizontal structures of a larger energy scale. The holographic interference structure is due to interference of recolliding trajectories (intra-cycle interference), along which the electron forward scatters by the atomic core. One of the trajectories is not significantly disturbed due to the rescattering (reference wave), while other paths are significantly disturbed (signal wave).  

The trajectories can be classified by the number of recollisions, which depend on the ionization time, i.e. the point in time where the electron is emitted into the continuum, or equivalently the final momentum $p_x$ of the electron along the laser field. Accordingly, in different regions of the PMD as a function of $p_x$, the number of the contributing trajectories is different, creating different topological structures. In the simplified simulation (Fig.~\ref{fig:ctmctra}(d)) three regions can be clearly identified: $|p_x|>\SI{0.75}{\atomicunit}$, $\SI{0.35}{\atomicunit}<|p_x|<\SI{0.75}{\atomicunit}$, and $|p_x|<\SI{0.35}{\atomicunit}$, also visible in in the full simulation, Fig.~\ref{fig:ctmctra} (c), and in the experimental results, Fig.~\ref{fig:O2vsXe} (more clearly in the lineouts). For $|p_x|>\SI{0.75}{\atomicunit}$ at given laser parameters, a single return to the atomic core exists (see the details in Sec.~\ref{topology}), and consequently, 2 types of recolliding trajectories, see Fig.~\ref{tra2}~(a),  with no significant rescattering and with one significant rescattering are possible. For smaller momenta ($\SI{0.35}{\atomicunit}<|p_x|<\SI{0.75}{\atomicunit}$), the trajectories have three recollisions. Therefore, in this case  8 types of  trajectories exist [with no significant rescattering, or one (taking place either at 1st, 2nd, or at 3rd recollision), two (at 1st and 2nd, or at 1st and 3rd, or at 2st and 3rd recollisions), or three significant rescatterings, depending on the initial transverse momentum], see Fig.~\ref{tra2}~(a)-(d). The smallest $p_x$ cutoff  for the trajectories with a single and three recollisions are  shown in Fig.~\ref{fig:ctmctra}~(b) in dependence on the laser field strength.

The initial momentum space, and the weight of the contribution of these recolliding trajectories are quite different. The largest contribution is from those with none and one significant rescattering. The initial momentum space of other trajectories is very small. This can be deduced from Fig.~\ref{tra2}~(e), which shows the relation between the initial, $p_{zi}$, and the final transverse momentum, $p_{z}$, at $p_x=\SI{0.5}{\atomicunit}$. The color of the line indicates the type of the trajectory, which are visualized in Fig.~\ref{tra2}~(a)-(d). The contributing initial momentum space  of each trajectory can be estimated by the $\delta p_{zi}$ region for the given $\delta p_{z}$, which is inversely proportional to the slope of the curves in Fig.~\ref{tra2}~(e). As Fig.~\ref{tra2}~(e) shows, the type of pair trajectories in each of panels (a)-(d) is interchanged with continuous variation of $p_{zi}$. For momenta smaller than $|p_x|<\SI{0.35}{\atomicunit}$, the electron revisits the atomic core more than three times, therefore, multiple rescatterings and more types of trajectories are possible.

\subsection{Topological structures in the PMD with respect to the longitudinal momentum}\label{topology}

We connect the cutoffs in $p_x$ momentum, determining the number of recollisions, with the regions where the topology of the interference pattern is unchanged, and define them in terms of the laser parameters. The cutoffs in  $p_x$ correspond to the slow recollision condition, when the longitudinal velocity at the recollision vanishes, $\dot{x}(\phi_r)=0$ (compare Fig.~\ref{fig:ctmctra}~(a)). 
The slow recollision can take place at one of the returns to the parent ion.
The ionization phase leading to a slow recollision at the $n$-th return can be approximated in a plane wave case (see Appendix~\ref{appendix:cutoff}): 
\begin{eqnarray}
	\phi_i(n) \approx 2/\left[(2n+1)\pi\right],
\end{eqnarray}
from which the cutoff momenta are found:
\begin{eqnarray}
p_x^{\text{cutoff}}(n) & = & \frac{E_0}{\omega}\sin{\left[\phi_i(n)\right]}\label{pxcutoff_}\\
p_x^{\text{cutoff}}(1^{\text{st}}) & \approx & 0.211  \frac{E_0}{\omega}, \label{pxcutoff1_}\\
p_x^{\text{cutoff}}(2^{\text{nd}}) & \approx & 0.127  \frac{E_0}{\omega},\label{pxcutoff2_}
\label{pxcutoffest}
\end{eqnarray}
with the laser field amplitude $E_0$, and the frequency $\omega$. These equations are illustrated in Fig.~\ref{fig:ctmctra}~(b) and can be used to classify interfering trajectories in strong-field holography. In the region $p_x>p_x^{\text{cutoff}}(1^{\text{st}})$, a single recollision exists,  at $p_x^{\text{cutoff}}(2^{\text{nd}})<p_x<p_x^{\text{cutoff}}(1^{\text{st}})$ three recollisions, and so on. Accordingly, the borders of the topologically uniform regions in the holographic interference pattern are  $p_x=p_x^{\text{cutoff}}(1^{\text{st}})$, $p_x=p_x^{\text{cutoff}}(2^{\text{nd}})$, and so on.

\begin{figure}[b]
	\centering
	\includegraphics[width=0.9\columnwidth]{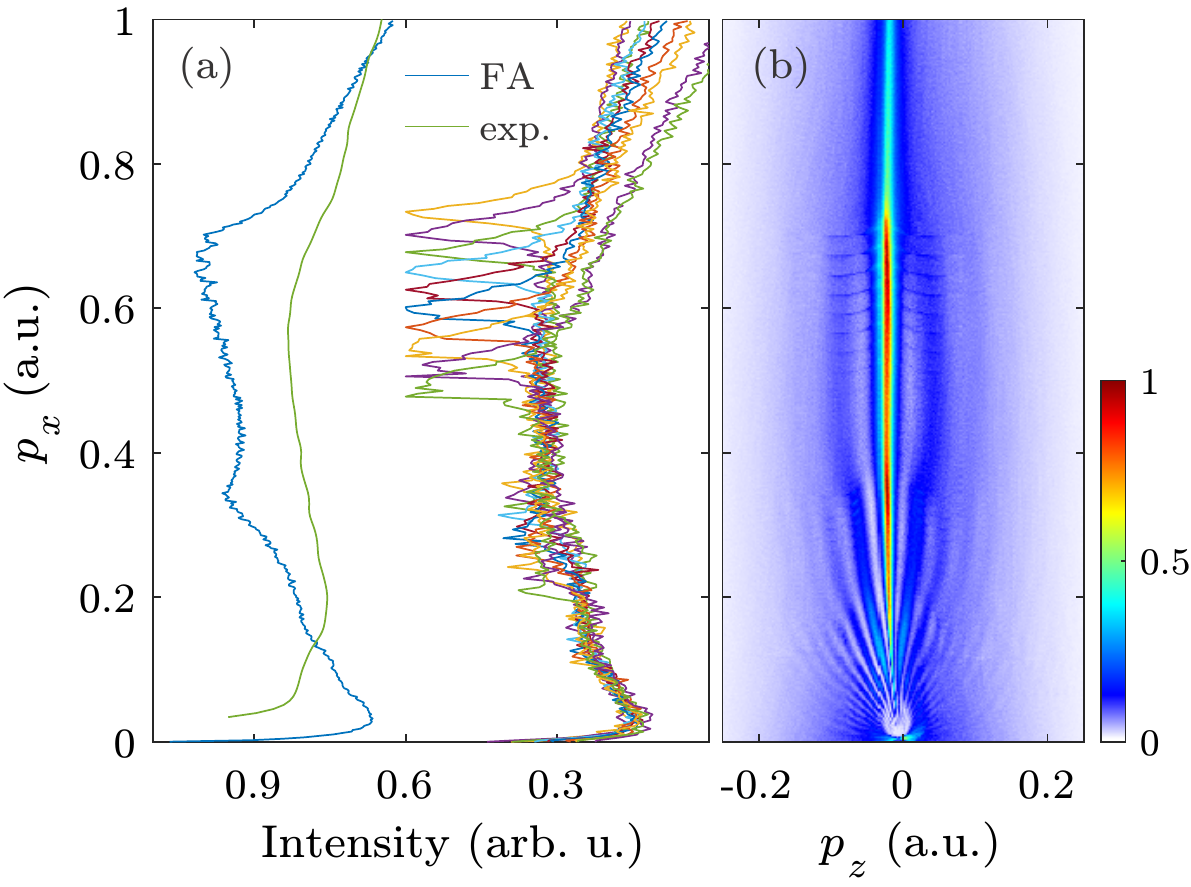}
	\caption{(a) Longitudinal momentum distribution.  The distributions for 10 sampling intensities are shown as overlapping multicolor lines.  The peaks for each intensity  correspond to the 1$^\text{th}$ transitions points (cutoff momenta)  of the holography pattern for the given intensity. The distribution of the total focal averaged spectrum from (b) can be found in (a) as blue line and the distribution from the experimental data as green line. The knee position of the $p_x$-distribution at $p_x \approx \SI{0.75}{\atomicunit}$ corresponds to the transition border of the topological structures for the intensity very close to the peak intensity.}
	\label{ddd}
\end{figure}

The model simulation, shown in Fig.~\ref{fig:ctmctra}(d) includes the two main types of the intra-cycle trajectories [blue and red in Fig.~\ref{tra2} (a) and (e)]: the  trajectory without significant rescattering  and the trajectory with the most significant rescattering (for $|p_x|>\SI{0.75}{\atomicunit}$ it takes place at the single recollision; for $\SI{0.35}{\atomicunit}<|p_x|<\SI{0.75}{\atomicunit}$  at the third recollision). 
When all trajectories are included, see Fig.~\ref{fig:ctmctra}(b), the momentum distribution becomes more smooth. However, the main features of the interference structure is already given by the spectra based on the two main trajectories. The PMD reveals an interference structure with a middle lobe and with side wings, that qualitatively were already known from the nonrelativistic regime \cite{Huismans_2011}. The nondipole effects shift the holography pattern to the negative $p_z$ direction breaking the symmetry with respect to $p_z$. The effect of the Coulomb field is to squeeze the interference structure in the transverse plane. 
The topology of the interference pattern changes at $p_x\approx \SI{0.35}{\atomicunit}$ and $p_x\approx \SI{0.75}{\atomicunit}$, due to the change of the number of recolliding trajectories. The cutoff value of $p_x$ at respective changes corresponds to the slow recollision conditions. Within each band of $p_x$, the $p_z$-positions of the sidelobes continuously vary with $p_x$, while discontinuity arises at the transition points of the bands.
 
The distribution of the intensities in the focal volume affects the transition points in $p_x$ between the regions with unchanged topology in the holographic interference pattern. 
These transition points can be more quantitatively identified in the longitudinal momentum distribution integrated over the transverse $p_z$ components, see Fig.~\ref{ddd}(a). We compare the experimental distribution with the QTMC simulation including the focal volume averaging and with the distributions corresponding to each intensity in the focal volume. The $p_x$-distribution for a single intensity has a peak corresponding to the end of the sidelobes of the holographic interference pattern. With larger intensity this peak moves to larger $p_x$. The simulated focal volume averaged $p_x$-distribution exhibits a dominant and from one side sharp maximum at $p_x \approx \SI{0.675}{\atomicunit}$, which rolls off to the knee at $p_x \approx \SI{0.75}{\atomicunit}$, followed by a smooth flat behavior. The experimental distribution does not have a sharp maximum, but shows a knee at the same $p_x \approx \SI{0.75}{\atomicunit}$. The knee position corresponds to the transition border of the topological structures for the intensity very close to the peak intensity. Although the focal volume of the intensities close to the peak intensity is small, along with the corresponding contribution in the PMD, nevertheless, the end of the sidelobes of the peak intensity, which is correlated with the transition region of the topological structures, can be visible in PMD due to the largest shift in $p_x$.  Thus, we conclude that the knee position can be used to determine the laser peak intensity according to Eq.~(\ref{pxcutoff_}). For the applied parameters, the cutoff $p_x=\SI{0.75}{\atomicunit}$ corresponds to the laser peak field $E_0=\SI{0.0528}{\atomicunit}$, i.e. laser intensity of $I_0=  \SI{9.8e13}{\watt\per\square\centi\meter}$.

\subsection{Nondipole effects}

Generally, the  nondipole effects disturb the recollision physics when the relativistic recollision parameter $\Gamma_R$ is large \cite{Palaniyappan_2006a}: 
\begin{eqnarray}
\Gamma_R\equiv \sqrt{\frac{U_p^3I_p}{2c^4\omega^2}},
\end{eqnarray}
with the electron ponderomotive potential $U_p=E_0^2/(4\omega^2)$. At $\Gamma_R=1$ the electron typical drift momentum in the laser propagation direction $p_z=U_p/c$ equals the momentum spread $\Delta_\bot$ of the tunneled electron wave packet transverse to the electric field, $\Delta_\bot=\sqrt{E_0/\sqrt{2I_p}}$ \cite{Popov_2004u} (or the size of the photoelectron wave packet spreading  at the recollision moment equals the relativistic drift distance). 
Although the relativistic recollision parameter is rather small in our experiment, $\Gamma_R\approx \SI{7.4e-3}{}$, the photoelectron momentum resolution in our experiment is sufficiently high to resolve the nondipole effect on the holography pattern of the order of $\delta p_z\sim \sqrt{\Gamma_R}\Delta_\bot=U_p/c\sim \SI{2e-2}{\atomicunit}$.

The nondipole signature in the holographic interference pattern is the asymmetry with respect to the laser propagation direction: a nonuniform shift of the momentum distribution along the laser propagation direction, already noted in \cite{Chelkowski_2015,Ivanov_2016,Brennecke_2018}. The shift fades out for vanishing longitudinal momenta. It is negative and increases in absolute value with the increase of $p_x$ (the right sidelobes are slightly stronger than the left ones for momenta $p_x>\SI{0.75}{\atomicunit}$). However, this behavior is reversed in the case of larger $p_x$. For $p_x>\SI{1.75}{\atomicunit}$, the shift of the interference structure is fully in positive $p_z$ direction (Fig.~\ref{fig:O2vsXe}). 

\subsubsection{The nondipole shift of the main lobe }

We study the scaling laws for the nondipole characteristic features of the holographic interference pattern. 
The position of the lobes of the holography structure is determined by the phase difference of the direct (trajectory 1) and the rescattered (trajectory 2) trajectories, which we estimate  using Eq.~(\ref{phase}). 
We find for the phase difference (see appendix~\ref{appendix2} for details):
\begin{eqnarray}
\Delta\varphi \approx  \left(\delta p_{z\,2}^C-\delta p_{z\,1}^C\right)\left[\delta p_{z\,2}^C+\delta p_{z\,1}^C-2(p_z+\overline{T}_z) \right] \frac{\phi_r-\phi_i}{2\omega},\nonumber\\
\label{deltaphi10}
\end{eqnarray}
with the average drift momentum during the recollision process
\begin{eqnarray}
\overline{T}_z &\equiv& \frac{1}{\phi_r-\phi_i} \int^{\phi_r}_{\phi_i} T_{z}(\phi) \,d\phi \\
 &=& \frac{1}{c (\phi_r-\phi_i)} \int^{\phi_r}_{\phi_i} \left[p_x A(\phi)+\frac{A^2(\phi)}{2} \right]\,d\phi \,.
\end{eqnarray}
Here, $\phi_i$ and $\phi_r$ denote the ionization and the recollision phase, respectively, and $\delta {p_{z}^C}_{1,2}$ are the Coulomb momentum transfer upon recollision for the trajectory 1 and 2, respectively. 

\begin{figure}[b]
	\centering
	\includegraphics[width=0.85\columnwidth,trim=0.2cm 0.1cm 0cm 0.0cm,clip]{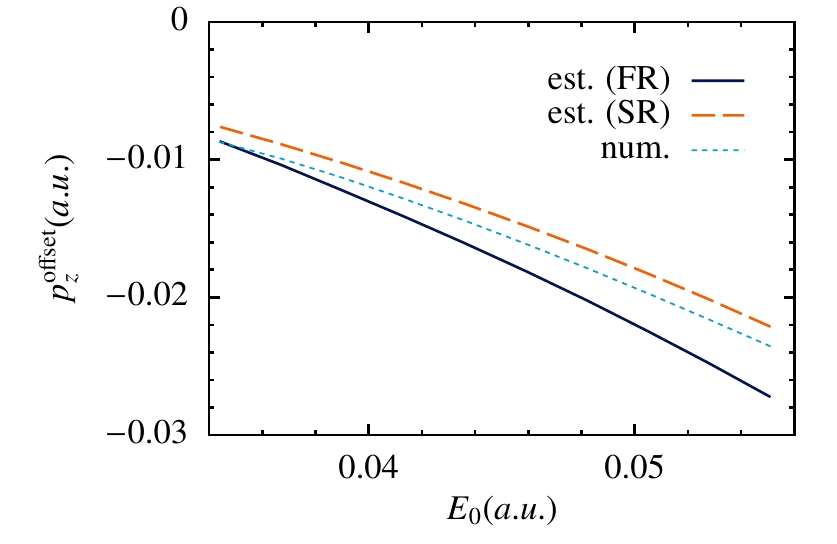}
	\caption{Nondipole shift of the PMD ridge  at $p_x = \SI{0.5}{\atomicunit}$ vs the laser field strength: The analytical prediction of Eq.~(\ref{ridge}), with $\phi_r$ corresponding to the first, fast recollision (FR) (solid), and corresponding to the  the second, slow recollision (SR) (long-dashed); The   numerically found classical trajectory that end at the ridge (short-dashed). } 
	\label{offsets}
\end{figure}

We can show that the main lobe of the quantum interference pattern ($\Delta\varphi=0$) coincides with the position of the sharp ridge due to Coulomb focusing \cite{Brennecke_2019,Danek_2018,Danek_2018c}. The electrons undergoing a single recollision and starting with $p_{zi}=-\overline{p}_{zd}-\delta p_z^C$ ($\overline{p}_{zd}$ is the average of the $p_{zd}(\phi)$ during the rescattering process), have the same recollision impact parameter as in the dipole case (the same $\delta p_{z}^C$), and end up on the ridge with 
\begin{eqnarray}
p_z^{(0)}=-\overline{T}_z.
\label{ridge}
\end{eqnarray}
Each  pair of trajectories contributing to the ridge which have $p_{zi\,1,2}=-\overline{p}_{zd}\pm |\delta p_z^C|$  with the same $|\delta p_z^C|$ interfere constructively with $\Delta \varphi=0$, creating the main lobe of the holographic interference pattern. This follows from Eqs.~(\ref{deltaphi10}) and (\ref{ridge}), and the condition $\delta p_{z\,1}^C+\delta p_{z\,2}^C=0$, fulfilled for these trajectories. 

In the case of multiple recollisions (at small values of the final $|p_x|$), the ridge position is closer to zero: $p_z^{(0)}=-\overline{T}_z+\tilde{p}_C$, where $\tilde{p}_C$ is positive and determined by multiple recollisions, see Eq.~(53) in \cite{Danek_2018}. We can prove that again the trajectories contributing to the ridge interfere constructively. For instance, we take two trajectories with $p_{yi\,1}=-p_{yi\,2}$ and $p_{zi\,1,2}=0$, which contribute to the ridge ($p_y=0$, $p_z=-\overline{T}_z+\tilde{p}_C$) and have $\delta p_{z\,1}^C=\delta p_{z\,2}^C$, yielding $\Delta \varphi =0$ according to Eq.~({\ref{deltaphi10}}). In this way the remarkable relation between the main lobe of the quantum interference pattern and the PMD ridge due to Coulomb focusing is confirmed. 		       		 		          
        		 		       		 		          
The negative offset given by Eq.~(\ref{ridge})  depends on the laser field  acting on the electron during the excursion in the continuum, taking place between the ionization and the recollision. The position of the ridge with respect to the laser electric field intensity at $p_x = \SI{0.5}{\atomicunit}$ is shown in Fig.~\ref{offsets}. We can see that at low intensities the numerically found position approaches to the result of Eq.~(\ref{ridge}), with $\phi_r$ corresponding to the first fast recollision. As the intensity increases, the numerical solution approaches the estimation for $\phi_r$ corresponding to the first slow recollision, which is shifting closer to the parent ion and starts gaining on importance.

Both, the ridge and the main lobe in QTMC demonstrate the same nonuniform dependence of the nondipole $p_z$-shift on the longitudinal momentum (Fig.~\ref{ddd}). In the case of a single recollision (for $p_x\gtrsim \SI{0.75}{\atomicunit}$ in the case of the applied laser parameters) the momentum of the main lobe is determined by Eq.~(\ref{ridge}). When the final  momentum is large such that rescattering is negligible, $|p_x|>\SI{1.75}{\atomicunit}$, the ridge shift is in the positive $z$-direction: $p_{z}^{(0)}\approx A(t_i)^2/2c$. For smaller $p_x$, it is negative.  The largest negative shift can be estimated $p_{z}^{(0)}\approx - U_p/c$.
For low $p_x<0.75$ multiple recollisions take place, and consequently, the average drift momentum decreases for increasing  order of recollision, and the negative shift of the ridge  declines \cite{Danek_2018}.

\subsubsection{Effect of the focal volume averaging for the main lobe shift}

\begin{figure}
	\centering
	\includegraphics[width=\columnwidth]{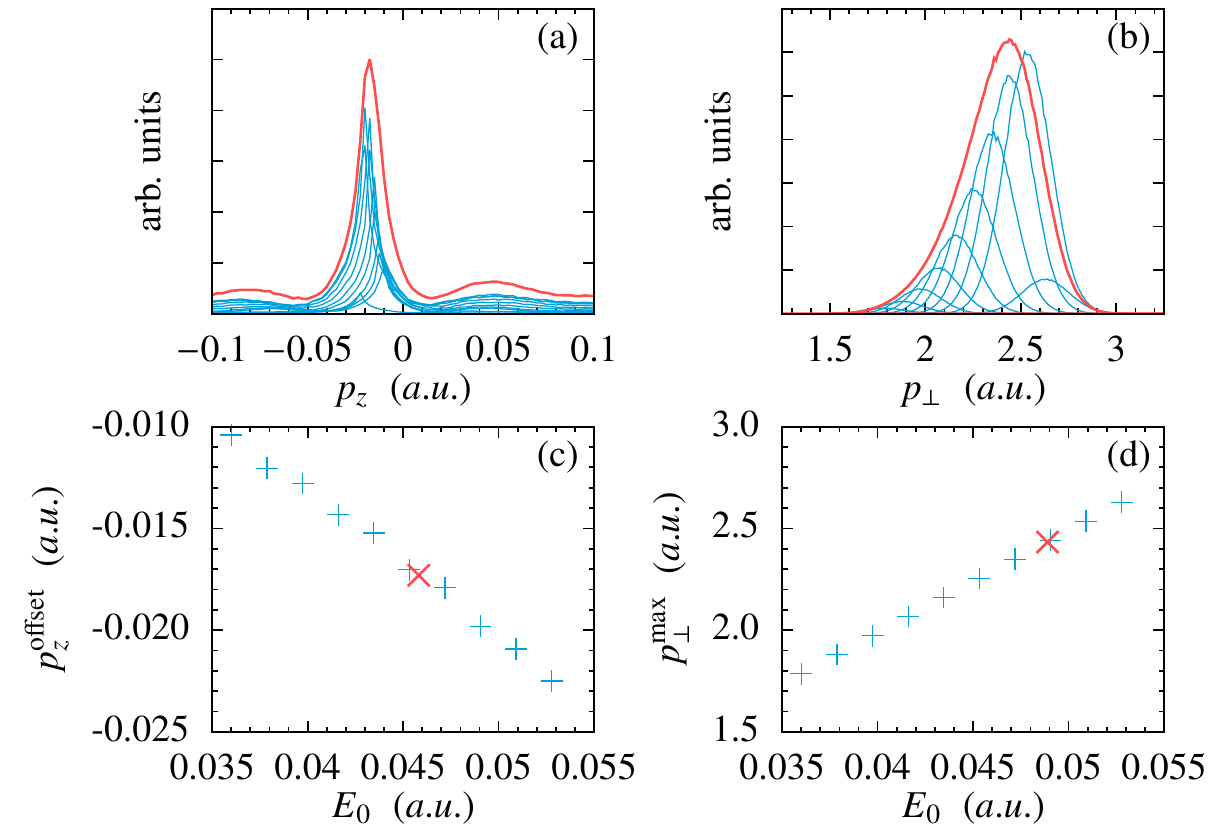}
	\caption{Focal volume averaging and the measured effective intensity: (a) for the ridge in the transverse PMD in linearly polarized laser field; (b) for the PMD radius in the polarization plane in a circularly polarized laser field. The distributions at each laser intensity are shown in blue. The position of the maxima in (a) and (b) are shown in (c) and (d), respectively. The focal averaged  intensity  deduced from the ridge in the case of linear polarization [red cross in (c)] is different from that deduced from the  case of a circular polarization [red cross in (d)].}
	\label{leFA}
\end{figure} 

The position of the ridge depends on the laser intensity and is disturbed because of focal volume averaging. We address this issue in Fig.~\ref{leFA}. We calculate the focal volume averaged momentum distribution of the ridge, from which the focal averaged intensity can be deduced (i.e. a single intensity which produces the same peak as the focal averaged ridge distribution).

With the same procedure we have also analyzed the focal volume averaging of the momentum distribution ring in the case of a circularly polarized laser field. We show that the focal averaging influences the position of the ridge (in the case of linear polarization) differently than the radius of the momentum distribution ring (in the case of circular polarization). Our conclusion is that the focal averaged intensity extracted from the circular polarization case \cite{Alnaser_2004,Smeenk_2011oe} cannot be applied to correctly reproduce the size of the holographic pattern in the PMD for linear polarization.

The nondipole momentum shift of the main lobe of the holography pattern can be employed for calibration of the focal volume averaged laser intensity. For the applied parameters, the main lobe position is $p_z=\SI{-0.009}{\atomicunit}$, at $p_x=\SI{0.5}{\atomicunit}$, which according to Fig.~\ref{offsets} gives the focal volume averaged field ${\overline E} = \SI{0.032}{\atomicunit}$, corresponding to the focal volume averaged laser intensity of ${\overline I}=  \SI{3.5e13}{\watt\per\square\centi\meter}$.

\subsubsection{Sidelobes}

The position of the sidelobes in the interference pattern is defined by the phase difference via
$\Delta\phi= 2\pi n$, with an integer~$n$. For rather large $n$ we can approximate  $\delta p_{z\,1}\approx 0$ for the direct trajectory, and $\delta p_{z\,2}\approx p_z$ for the rescattered trajectory. In that case Eq.~(\ref{deltaphi10}) yields
\begin{eqnarray}
\Delta\phi \approx  \left[\frac{p_z^2}{2} +p_z \overline{T}_z \right]  \frac{\phi_r-\phi_i}{\omega}=2\pi n.
\label{deltaphi3}
\end{eqnarray}
In the dipole limit ($\overline{T}_z=0$) the sidelobe positions are 
\begin{eqnarray}
p_{z\,dip}^{(n)}=\pm\sqrt{\frac{4\pi n\omega}{\phi_r-\phi_i}},
\label{sidelobedip}
\end{eqnarray}
and nondipole corrections shift the sidelobes slightly: $p_{z}^{(n)}\approx p_{z\,dip}^{(n)}+\delta p_z$, with $\delta p_z\ll p_{z\,dip}^{(n)}$. From Eq.~(\ref{sidelobedip}), $\delta p_z=-\overline{T}_z$, and the sidelobe positions in the nondipole case are:
\begin{eqnarray}
p_{z }^{(n)}=\pm\sqrt{\frac{4\pi n\omega}{\phi_r-\phi_i}}-\overline{T}_z.
\label{sidelobe}
\end{eqnarray}
The latter indicates that the holography pattern in the nondipole regime is shifted as a whole for a fixed $p_x$, in the direction opposite to the laser propagation direction. However, the shift is not uniform and depends on $p_x$, similar to the ridge. Qualitatively we can estimate the rescattering phase difference as $(\phi_r-\phi_i)\sim 2\pi\delta $, assuming the rescattering time is a $\delta$-fraction of the laser period.
The distance between, e.g., the main and the 2nd lobes in momentum space is  $|p^{(1)}_z-p^{(0)}_z|\approx \sqrt{2\omega/\delta}$, which applies in the region of a single rescattering $p_x>\SI{0.75}{\atomicunit}$. For $p_x<\SI{0.75}{\atomicunit}$ multiple recollisions take place which increases the effective recollision time, consequently, decreasing the separation of the lobes.

Equation (\ref{sidelobe}) explains also why the left interference lobe is weaker than the right one. In fact, for the direct electrons $p_{zi}^{(n)}=p_z^{(n)}$, and the initial transverse momentum for the electron contributing to the left side-lobes is larger than that of the right ones of the same $n^{th}$-order. Therefore,  the probability of the left side-lobes $W^{(n-)}$ are suppressed compared to the right-lobes $W^{(n+)}$ due to a smaller tunneling probability: 
\begin{eqnarray}
\frac{W^{(n-)}}{W^{(n+)}}&=&\exp\left( -\frac{p_{z}^{(n-)2}-p_{z}^{(n+)2}}{\Delta_\bot^2}\right)=\exp\left( -\frac{4|p_{z,\,dip}^{(n)}|\overline{T}_z}{\Delta_\bot^2}\right)\nonumber \\
&&\approx \exp\left( -2\sqrt{\frac{nI_p}{\omega\delta}}\frac{E_0}{c\omega}\right),
\end{eqnarray}
where the estimation $\overline{T}_z\approx U_p/c$ is used. For instance $W_{-1}/W_1\approx 0.75$ at given parameters. With these simple estimations all qualitative features of the interference structure can be reproduced, showing how the laser magnetic field interaction alters the holography image of the momentum distribution.

\subsection{The role of the accurate description of the quantum scattering phase}

\begin{figure}
	\centering
	\includegraphics[width=0.8\columnwidth]{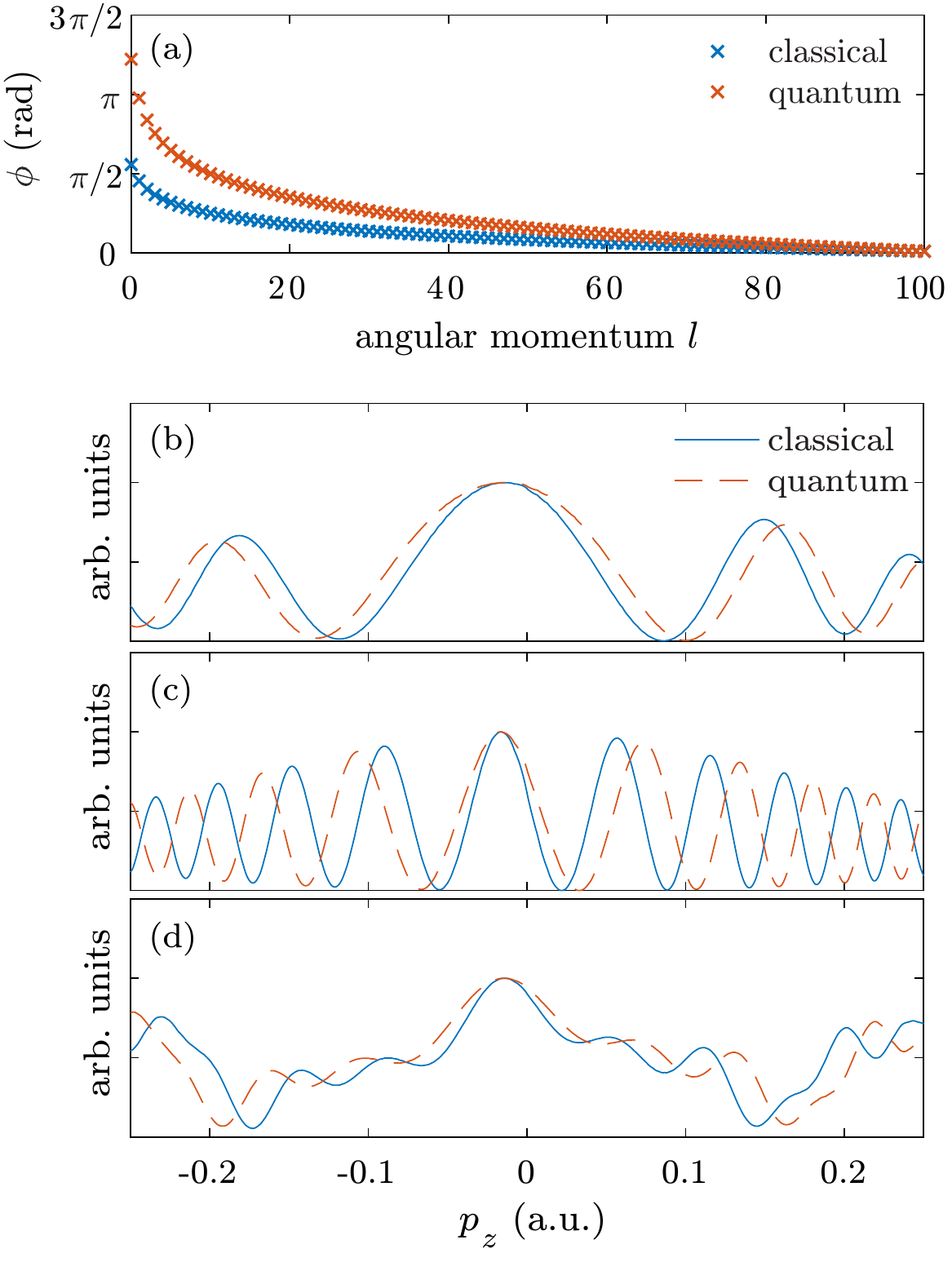}
	\caption{Comparison of quantum and quasiclassical recollision phases (a). The lineouts of the 2D calculation of PMD at $p_x = \SI{0.9}{\atomicunit}$ on panel (b) and $p_x = \SI{0.5}{\atomicunit}$ on panels (c) and (d) taking into account two (b) and (c) or all eight (d) interfering trajectories. The blue full lines represent the 2D calculation with the common classical SFA Coulomb phase and the the red dashed lines the results corrected by the quantum phase at each rescattering.}
	\label{PESphase}
\end{figure} 

\begin{figure*}
\centering
\includegraphics[width=0.85\textwidth]{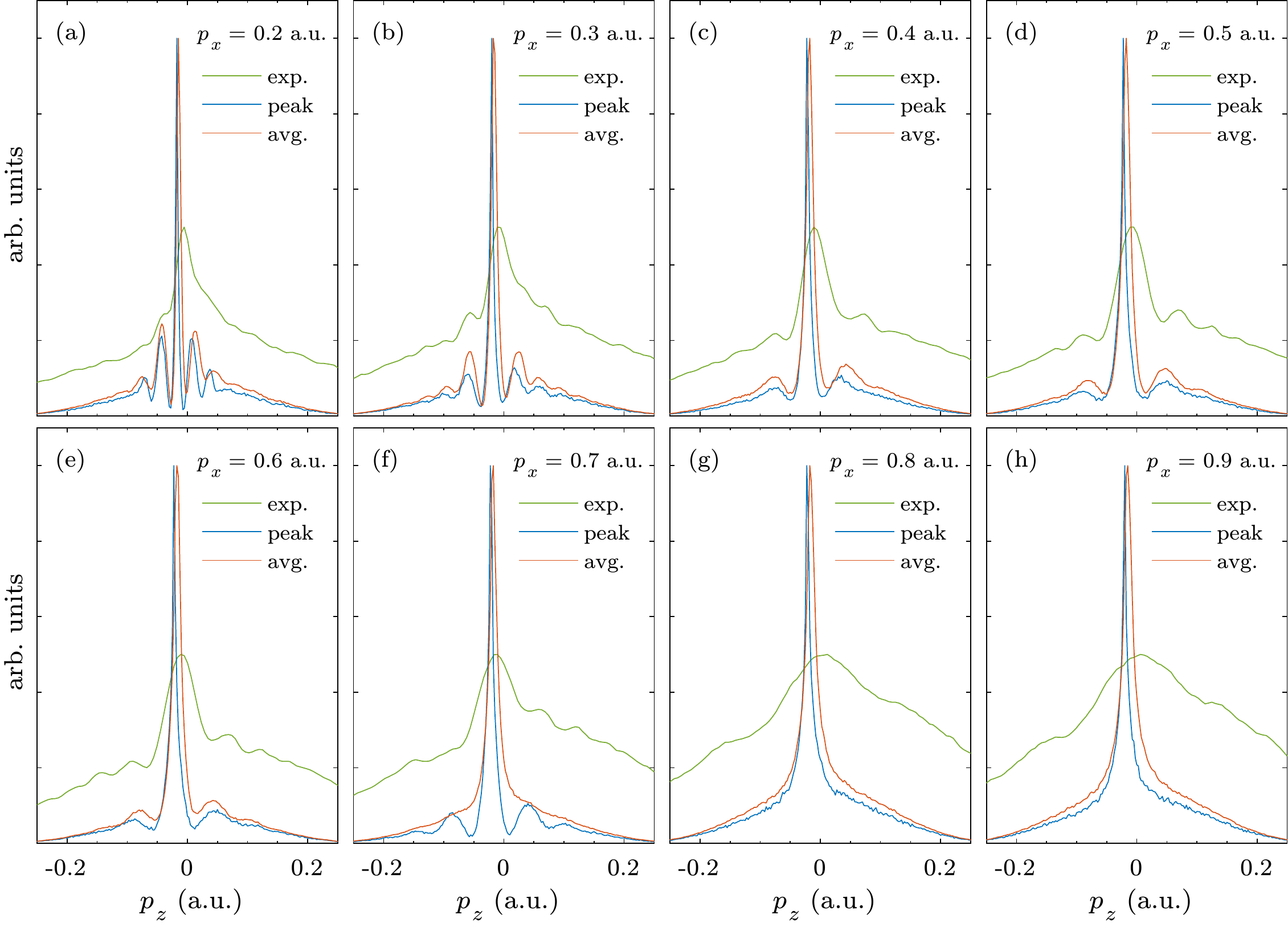}
	\caption{Comparison of the lineouts obtained from experiment (exp.) and QTMC calculation with a focal averaged intensity (avg.) and the peak intensity (peak) for different positions along the laser polarization axis $p_x$.}
	\label{lineouts_Xe}
\end{figure*} 

We observe small discrepancies between the PMD lineouts of the experiment and the QTMC simulations. The sidelobes from the QTMC simulations are closer to the main lobe than in the experiment (Fig.~\ref{lineouts_Xe}). We assume that the latter comes from the fact that in the QTMC simulates the rescattering phase is based on the quasiclassical approximation, which slightly deviates from the exact quantum scattering phase.
The quantum rescattering phase was rigorously calculated for several potentials \cite{Friedrich_Scattering}. The quantum phase acquired by an electron during scattering off the Coulomb potential is 
\begin{eqnarray}
\phi^C = arg\left[\Gamma(l+1+i\eta)\right],
\end{eqnarray}
where $\Gamma$ is the gamma function, $l$ is the quantum number of the orbital angular momentum,  and $\eta =  Z/p$ is the Sommerfeld parameter with the electron momentum $p$ and charge $Z$. 
On the other hand, in our quasiclassical simulation the Coulomb scattering phase is calculated analytically along the classical trajectory $\mathbf{r}(t)$ as 
\begin{eqnarray}
\phi^C_{class} = \int\limits_{-\infty}^{+\infty}\text{d} t\, Z/r(t), 
\end{eqnarray}
which after   extraction of the divergence   \cite{Shvetsov-Shilovski_2016} yields
\begin{eqnarray}
\phi^C_{class} = -\frac{Z}{\sqrt{2{\cal E} }} \ln{\left(1+2{\cal E} L^2\right)},
\end{eqnarray}
where $L$ is the total angular momentum and ${\cal E}$ is the energy of the incoming electron. The comparison of the quasicalssical scattering phase with the exact one shows that the difference between the phases can reach up to $\pi/2$ for small angular momenta, which may affect the interference pattern and positions of the fringes (Fig.~\ref{PESphase} (a)). 
We analyze the role of the quantum phase on the interference pattern and focus on two slices at $p_x = \SI{0.8}{\atomicunit}$ and $p_x = \SI{0.5}{\atomicunit}$, and the cases of interference of two or all (eight) trajectories (Fig.~\ref{PESphase} (b)-(d)). For these values of $p_x$, all recollisions are fast, taking place at the maximal speed. Therefore, we can assume that the recollision resembles the field free case  and apply the phase correction $ \phi^C - \phi^C_{class} $. As we can see, in all cases the correction of the scattering phase leads to widening of the interference pattern. 

We also analyzed numerically the role of the under-the-barrier dynamics for the phase difference of the direct and rescattered trajectories, with a conclusion that its effect on the phase difference is negligible.

\section{Conclusion}
\label{conclusion}

In our experiment we have resolved the holographic interference structures in the PMD with a momentum precision of \SI{1e-2}{\atomicunit} which enables us to discern the signatures of the nondipole interaction with the laser field at nonrelativistic laser intensities. We show that the competing effect of the laser magnetic field induced drift in beam propagation direction and Coulomb focusing explain the longitudinal momentum dependent shift of the holography pattern. We prove that the main lobe of the interference pattern coincides with the ridge described fully classical by Coulomb focusing. Its position in momentum along the propagation direction is positive at large longitudinal momenta, negative at intermediate and tends towards zero at low longitudinal momenta. We provide analytical estimates for the nondipole momentum shift of the holography interference pattern, as well as for the ratio of the intensities of the left to right sidelobes. 

We show that the focal averaging alters the position of the ridge from which the focal averaged intensity can be deduced. The latter is shown to deviate from the focal averaged intensity extracted from the radius of the momentum distribution ring in the case of circular polarization. 
Consequently, care has to be taken when using the focal averaged intensity read out from the circular polarization case for accurate predictions about the PMD's holographic interference pattern. 

We relate the change of the topological structure of the holography pattern as a function of the longitudinal momentum to the number of recolliding trajectories, and show that the transition points of the topological structure are very sensitive to the intensity. In particular, these transition points encode the peak intensity during the strong-field ionization process.

Finally, we explain the slight discrepancy in the position of the interference lobes between QTMC and the experiment by the deviation of the quasiclassical scattering phase from the exact quantum phase.

\appendix
\section{Derivation of the cutoff momenta}
\label{appendix:cutoff}
The phase at the slow recollision condition can be estimated from the longitudinal component of the laser driven trajectory (for the estimate we use a monochromatic laser field described by the vector potential $A(\phi) = E_0 / \omega \cos(\phi)$):
\begin{eqnarray}
x_0(\phi) & = & \frac{E_0}{\omega^2}\left[\cos{\phi}-\cos{\phi_i} + \left(\phi - \phi_i\right)\sin{\phi_i}\right] \nonumber \\ &&+ \frac{p_{xi}(\phi_i)}{\omega}\left(\phi - \phi_i\right) + x_i(\phi_i),
\label{x0}
\end{eqnarray}
where the trajectory evolves with the phase of the laser field $\phi$  and starts at ionization phase $\phi_i$. Here, $x_i(\phi_i)\approx  - I_p/(E_0\cos{\phi_i})$ is the tunnel exit,  and $p_{xi}(\phi_i)=Z\pi  E_0 \cos{\phi_i}{(2 I_p)^{-3/2}}$  is the  Coulomb momentum transfer to the phototelectron  at the tunnel exit \cite{Shvetsov-Shilovski_2009}, with the charge $Z$ of the atomic core, which we have included in the initial condition for simplicity. From the slow recollision conditions $\dot{x}_0(\phi_r)=0$, the phase of the $n^{th}$ slow recollision  can be derived
\begin{equation}
\phi_r(n) = -\phi_i(n) + \left(2n+1\right)\pi - \delta_n,
\label{urn}
\end{equation}
where $\delta_n \equiv \omega p_{xi}\left(\phi_i(n)\right)/(E_0\cos{\left(\phi_i(n)\right)})$ accounts for the Coulomb effects at the moment of ionization.
The ionization phase $\phi_i(n)$ leading to the $n^{\text{th}}$ slow recollision can be now determined from the general recollision condition $x_0 \left(\phi_r \right) = 0$. Using Eq.~(\ref{x0}) and the leading terms of the expansion over $\phi_i$ and $\delta_n$, one obtains:
\begin{equation}
\phi_i(n) \approx \frac{2\left(1+\frac{\gamma_K^2}{4}\right) - (2n+1)\pi^2 \frac{Z}{\kappa} \frac{E_0}{E_a}\gamma_K\left(1-\frac{1}{2n+1}\frac{Z}{\kappa} \frac{E_0}{E_a}\gamma_K\right)}{
(2n+1)\pi\left(1-\frac{2}{2n+1}\frac{Z}{\kappa} \frac{E_0}{E_a}\gamma_K\right)
}.
\label{phin}
\end{equation}
Here $\kappa = \sqrt{2 I_p}$ is the characteristic momentum of an electron from a target with ionization potential $I_p$, $E_a = (2I_p)^{(3/2)}$ the characteristic atomic field of the target and $\gamma_K = \omega \kappa / E_0$ the Keldysh parameter.
For the employed parameters ($\kappa\approx 1$\,a.u., $E_0 / E_a \ll 1$ and $\gamma_K \approx 0.4$) Eq.~(\ref{phin}) simplifies to 
\begin{eqnarray}
\phi_i(n) \approx 2/\left[(2n+1)\pi\right],
\label{phins}
\end{eqnarray} 
and from the latter, the cutoff momenta are found:
\begin{eqnarray}
p_x^{\text{cutoff}}(n) & = & (E_0/\omega)\sin{\left[\phi_i(n)\right]}\label{pxcutoff}\\
p_x^{\text{cutoff}}(1^{\text{st}}) & \approx & 0.211  \frac{E_0}{\omega}, \label{pxcutoff1}\\
p_x^{\text{cutoff}}(2^{\text{nd}}) & \approx & 0.127  \frac{E_0}{\omega}.\label{pxcutoff2}
\end{eqnarray}

\section{Derivation of the phase difference for direct and rescattered trajectories}\label{appendix2}

We start from Eq.~(\ref{phase}) and replace the last term using the equation for the kinetic energy evolution: 
\begin{eqnarray}
\frac{d\varepsilon}{dt}=-\textbf{v}\cdot \textbf{E}-\textbf{v}\cdot{\boldsymbol\nabla} V(\textbf{r})\,.
\end{eqnarray}
This allows integration by parts, yielding
\begin{eqnarray}
{\cal S}^{(l)}(\phi,\mathbf{p})&\approx &  - \int^{\phi_f}_{\phi}\frac{d\phi'}{\omega}\left\{ \left[\varepsilon+V(\textbf{r})\right] -\frac{\mathbf{r}\cdot{\boldsymbol\nabla } V(\textbf{r})}{1-\beta_z} \right\}+I_p(t_f-t)  ,\nonumber\\
\label{phase1}
\end{eqnarray}
where $ \beta_z=v_z/c$ is the Lorentz-factor, and the integration variable is changed from the time $t$ to the laser phase $\phi=\omega(t-z/c)$.
Then, the classical action in the order of $1/c$ will read
\begin{eqnarray}
{\cal S}^{(l)}(\phi,\mathbf{p})&\approx &  - \int^{\phi_f}_{\phi}d\phi'\left\{  \frac{\textbf{p}^2(\phi'))}{2} + V(\textbf{r}(\phi'))-\mathbf{r}(\phi')\cdot{\boldsymbol\nabla } V(\textbf{r}(\phi'))\right\}\nonumber\\&&+I_p(t_f-t) .
\label{phase1c}
\end{eqnarray}
Let us estimate the phase difference of the interfering trajectories in the nondipole case, which creates the holography pattern. The main phase difference arises during the electron dynamics between  the ionization and recollision, because the momenta of the direct and rescattering trajectories coincide after the recollision \cite{Huismans_2012}. Further, we assume that the Coulomb momentum transfer to the electron takes place at the recollision points, while the electron motion during the excursion is governed only by the laser field \cite{Maurer_2018}:
\begin{eqnarray}
p_x(\phi)&=&A(\phi)-A(\phi_i) \nonumber\\
p_y(\phi)&=&p_{yi}\label{pxyz}\\
p_z(\phi)&=&p_{zi}+p_{zd}(\phi) \nonumber\
\end{eqnarray}
with the drift momentum $p_{zd}(\phi)=[A(\phi)-A(\phi_i)]^2/(2c)$, and the initial momentum at the tunnel exit $\mathbf{p}_i$. As the $p_x$ and $p_y $ momentum components for the direct and rescattering trajectories are the same, the phase difference is derived:
\begin{eqnarray}
\Delta\varphi \approx\int^{\phi_r}_{\phi_i}\frac{d\phi}{\omega} \, \left[\frac{p_{z2}^2 (\phi) }{2}-\frac{p_{z1}^2(\phi)}{2}\right] ,
\label{deltaphi1}
\end{eqnarray}
where $\phi_1,\,\phi_r$ are the ionization and the recollision phases, and $p_{z1}(\phi)\,,p_{z2}(\phi)$ are the direct and rescattered electron momenta given by Eq.~(\ref{pxyz}). We can express Eq.~(\ref{pxyz}) via the electron final momentum $p_z=p_{zi}+A^2(\phi_i)/(2c)+\delta p_z^C$,  which is the same for the direct and rescattered electrons:
\begin{eqnarray}
p_{z\,1,2}(\phi)= p_z +T_{z}(\phi)-\delta p_{z\,1,2}^C.
\label{pz12}
\end{eqnarray}
with the Coulomb momentum transfer at the recollision $\delta p_{z\,1,2}^C$, and
\begin{eqnarray}
T_z( \phi)\equiv \frac{1}{c}\left[p_x A(\phi)+\frac{A^2(\phi)}{2}\right ],
\end{eqnarray}
is the contracted drift momentum along the laser propagation direction, with the asymptotic momentum $p_x$. With Eqs.~(\ref{deltaphi1})-(\ref{pz12}) the phase difference is derived:
\begin{eqnarray}
\Delta\varphi \approx  \left(\delta p_{z\,2}^C-\delta p_{z\,1}^C\right)\left[\delta p_{z\,2}^C+\delta p_{z\,1}^C-2(p_z+\overline{T}_z) \right] \frac{\phi_r-\phi_i}{2\omega},\nonumber\\
\label{deltaphi2}
\end{eqnarray}
with the average contracted drift momentum during the recollision process
\begin{eqnarray}
\overline{T}_z\equiv \int^{\phi_r}_{\phi_i} \frac{d\phi}{\phi_r-\phi_i}  T_{z}(\phi).
\end{eqnarray}

\begin{acknowledgments}
This research was supported by the NCCR MUST, funded by the Swiss National Science Foundation and by the ERC advanced grant ERC-2012-ADG 20120216 within the seventh framework programme of the European Union. B. W. was supported by an ETH Research Grant ETH-11~15-1.
\end{acknowledgments}

\bibliography{holo}

\begin{thebibliography}{82}%
\makeatletter
\providecommand \@ifxundefined [1]{%
 \@ifx{#1\undefined}
}%
\providecommand \@ifnum [1]{%
 \ifnum #1\expandafter \@firstoftwo
 \else \expandafter \@secondoftwo
 \fi
}%
\providecommand \@ifx [1]{%
 \ifx #1\expandafter \@firstoftwo
 \else \expandafter \@secondoftwo
 \fi
}%
\providecommand \natexlab [1]{#1}%
\providecommand \enquote  [1]{``#1''}%
\providecommand \bibnamefont  [1]{#1}%
\providecommand \bibfnamefont [1]{#1}%
\providecommand \citenamefont [1]{#1}%
\providecommand \href@noop [0]{\@secondoftwo}%
\providecommand \href [0]{\begingroup \@sanitize@url \@href}%
\providecommand \@href[1]{\@@startlink{#1}\@@href}%
\providecommand \@@href[1]{\endgroup#1\@@endlink}%
\providecommand \@sanitize@url [0]{\catcode `\\12\catcode `\$12\catcode
  `\&12\catcode `\#12\catcode `\^12\catcode `\_12\catcode `\%12\relax}%
\providecommand \@@startlink[1]{}%
\providecommand \@@endlink[0]{}%
\providecommand \url  [0]{\begingroup\@sanitize@url \@url }%
\providecommand \@url [1]{\endgroup\@href {#1}{\urlprefix }}%
\providecommand \urlprefix  [0]{URL }%
\providecommand \Eprint [0]{\href }%
\providecommand \doibase [0]{https://doi.org/}%
\providecommand \selectlanguage [0]{\@gobble}%
\providecommand \bibinfo  [0]{\@secondoftwo}%
\providecommand \bibfield  [0]{\@secondoftwo}%
\providecommand \translation [1]{[#1]}%
\providecommand \BibitemOpen [0]{}%
\providecommand \bibitemStop [0]{}%
\providecommand \bibitemNoStop [0]{.\EOS\space}%
\providecommand \EOS [0]{\spacefactor3000\relax}%
\providecommand \BibitemShut  [1]{\csname bibitem#1\endcsname}%
\let\auto@bib@innerbib\@empty
\bibitem [{\citenamefont {Huismans}\ \emph {et~al.}(2011)\citenamefont
  {Huismans}, \citenamefont {Rouz{\'{e}}e}, \citenamefont {Gijsbertsen},
  \citenamefont {Jungmann}, \citenamefont {Smolkowska}, \citenamefont {Logman},
  \citenamefont {L{\'{e}}pine}, \citenamefont {Cauchy}, \citenamefont {Zamith},
  \citenamefont {Marchenko}, \citenamefont {Bakker}, \citenamefont {Berden},
  \citenamefont {Redlich}, \citenamefont {van~der Meer}, \citenamefont
  {Muller}, \citenamefont {Vermin}, \citenamefont {Schafer}, \citenamefont
  {Spanner}, \citenamefont {Ivanov}, \citenamefont {Smirnova}, \citenamefont
  {Bauer}, \citenamefont {Popruzhenko},\ and\ \citenamefont
  {Vrakking}}]{Huismans_2011}%
  \BibitemOpen
  \bibfield  {author} {\bibinfo {author} {\bibfnamefont {Y.}~\bibnamefont
  {Huismans}}, \bibinfo {author} {\bibfnamefont {A.}~\bibnamefont
  {Rouz{\'{e}}e}}, \bibinfo {author} {\bibfnamefont {A.}~\bibnamefont
  {Gijsbertsen}}, \bibinfo {author} {\bibfnamefont {J.~H.}\ \bibnamefont
  {Jungmann}}, \bibinfo {author} {\bibfnamefont {A.~S.}\ \bibnamefont
  {Smolkowska}}, \bibinfo {author} {\bibfnamefont {P.~S. W.~M.}\ \bibnamefont
  {Logman}}, \bibinfo {author} {\bibfnamefont {F.}~\bibnamefont
  {L{\'{e}}pine}}, \bibinfo {author} {\bibfnamefont {C.}~\bibnamefont
  {Cauchy}}, \bibinfo {author} {\bibfnamefont {S.}~\bibnamefont {Zamith}},
  \bibinfo {author} {\bibfnamefont {T.}~\bibnamefont {Marchenko}}, \bibinfo
  {author} {\bibfnamefont {J.~M.}\ \bibnamefont {Bakker}}, \bibinfo {author}
  {\bibfnamefont {G.}~\bibnamefont {Berden}}, \bibinfo {author} {\bibfnamefont
  {B.}~\bibnamefont {Redlich}}, \bibinfo {author} {\bibfnamefont {A.~F.~G.}\
  \bibnamefont {van~der Meer}}, \bibinfo {author} {\bibfnamefont {H.~G.}\
  \bibnamefont {Muller}}, \bibinfo {author} {\bibfnamefont {W.}~\bibnamefont
  {Vermin}}, \bibinfo {author} {\bibfnamefont {K.~J.}\ \bibnamefont {Schafer}},
  \bibinfo {author} {\bibfnamefont {M.}~\bibnamefont {Spanner}}, \bibinfo
  {author} {\bibfnamefont {M.~Y.}\ \bibnamefont {Ivanov}}, \bibinfo {author}
  {\bibfnamefont {O.}~\bibnamefont {Smirnova}}, \bibinfo {author}
  {\bibfnamefont {D.}~\bibnamefont {Bauer}}, \bibinfo {author} {\bibfnamefont
  {S.~V.}\ \bibnamefont {Popruzhenko}},\ and\ \bibinfo {author} {\bibfnamefont
  {M.~J.~J.}\ \bibnamefont {Vrakking}},\ }\href
  {https://doi.org/10.1126/science.1198450} {\bibfield  {journal} {\bibinfo
  {journal} {Science}\ }\textbf {\bibinfo {volume} {331}},\ \bibinfo {pages}
  {61} (\bibinfo {year} {2011})}\BibitemShut {NoStop}%
\bibitem [{\citenamefont {Bian}\ \emph {et~al.}(2011)\citenamefont {Bian},
  \citenamefont {Huismans}, \citenamefont {Smirnova}, \citenamefont {Yuan},
  \citenamefont {Vrakking},\ and\ \citenamefont {Bandrauk}}]{Bian_2011}%
  \BibitemOpen
  \bibfield  {author} {\bibinfo {author} {\bibfnamefont {X.-B.}\ \bibnamefont
  {Bian}}, \bibinfo {author} {\bibfnamefont {Y.}~\bibnamefont {Huismans}},
  \bibinfo {author} {\bibfnamefont {O.}~\bibnamefont {Smirnova}}, \bibinfo
  {author} {\bibfnamefont {K.-J.}\ \bibnamefont {Yuan}}, \bibinfo {author}
  {\bibfnamefont {M.~J.~J.}\ \bibnamefont {Vrakking}},\ and\ \bibinfo {author}
  {\bibfnamefont {A.~D.}\ \bibnamefont {Bandrauk}},\ }\href@noop {} {\bibfield
  {journal} {\bibinfo  {journal} {Phys. Rev. A}\ }\textbf {\bibinfo {volume}
  {84}},\ \bibinfo {pages} {043420} (\bibinfo {year} {2011})}\BibitemShut
  {NoStop}%
\bibitem [{\citenamefont {Marchenko}\ \emph {et~al.}(2011)\citenamefont
  {Marchenko}, \citenamefont {Huismans}, \citenamefont {Schafer},\ and\
  \citenamefont {Vrakking}}]{Marchenko_2011}%
  \BibitemOpen
  \bibfield  {author} {\bibinfo {author} {\bibfnamefont {T.}~\bibnamefont
  {Marchenko}}, \bibinfo {author} {\bibfnamefont {Y.}~\bibnamefont {Huismans}},
  \bibinfo {author} {\bibfnamefont {K.~J.}\ \bibnamefont {Schafer}},\ and\
  \bibinfo {author} {\bibfnamefont {M.~J.~J.}\ \bibnamefont {Vrakking}},\
  }\href@noop {} {\bibfield  {journal} {\bibinfo  {journal} {Phys. Rev. A}\
  }\textbf {\bibinfo {volume} {84}},\ \bibinfo {pages} {053427} (\bibinfo
  {year} {2011})}\BibitemShut {NoStop}%
\bibitem [{\citenamefont {Huismans}\ \emph {et~al.}(2012)\citenamefont
  {Huismans}, \citenamefont {Gijsbertsen}, \citenamefont {Smolkowska},
  \citenamefont {Jungmann}, \citenamefont {Rouz{\'{e}}e}, \citenamefont
  {Logman}, \citenamefont {L{\'{e}}pine}, \citenamefont {Cauchy}, \citenamefont
  {Zamith}, \citenamefont {Marchenko}, \citenamefont {Bakker}, \citenamefont
  {Berden}, \citenamefont {Redlich}, \citenamefont {van~der Meer},
  \citenamefont {Ivanov}, \citenamefont {Yan}, \citenamefont {Bauer},
  \citenamefont {Smirnova},\ and\ \citenamefont {Vrakking}}]{Huismans_2012}%
  \BibitemOpen
  \bibfield  {author} {\bibinfo {author} {\bibfnamefont {Y.}~\bibnamefont
  {Huismans}}, \bibinfo {author} {\bibfnamefont {A.}~\bibnamefont
  {Gijsbertsen}}, \bibinfo {author} {\bibfnamefont {A.~S.}\ \bibnamefont
  {Smolkowska}}, \bibinfo {author} {\bibfnamefont {J.~H.}\ \bibnamefont
  {Jungmann}}, \bibinfo {author} {\bibfnamefont {A.}~\bibnamefont
  {Rouz{\'{e}}e}}, \bibinfo {author} {\bibfnamefont {P.~S. W.~M.}\ \bibnamefont
  {Logman}}, \bibinfo {author} {\bibfnamefont {F.}~\bibnamefont
  {L{\'{e}}pine}}, \bibinfo {author} {\bibfnamefont {C.}~\bibnamefont
  {Cauchy}}, \bibinfo {author} {\bibfnamefont {S.}~\bibnamefont {Zamith}},
  \bibinfo {author} {\bibfnamefont {T.}~\bibnamefont {Marchenko}}, \bibinfo
  {author} {\bibfnamefont {J.~M.}\ \bibnamefont {Bakker}}, \bibinfo {author}
  {\bibfnamefont {G.}~\bibnamefont {Berden}}, \bibinfo {author} {\bibfnamefont
  {B.}~\bibnamefont {Redlich}}, \bibinfo {author} {\bibfnamefont {A.~F.~G.}\
  \bibnamefont {van~der Meer}}, \bibinfo {author} {\bibfnamefont {M.~Y.}\
  \bibnamefont {Ivanov}}, \bibinfo {author} {\bibfnamefont {T.-M.}\
  \bibnamefont {Yan}}, \bibinfo {author} {\bibfnamefont {D.}~\bibnamefont
  {Bauer}}, \bibinfo {author} {\bibfnamefont {O.}~\bibnamefont {Smirnova}},\
  and\ \bibinfo {author} {\bibfnamefont {M.~J.~J.}\ \bibnamefont {Vrakking}},\
  }\href@noop {} {\bibfield  {journal} {\bibinfo  {journal} {Phys. Rev. Lett.}\
  }\textbf {\bibinfo {volume} {109}},\ \bibinfo {pages} {013002} (\bibinfo
  {year} {2012})}\BibitemShut {NoStop}%
\bibitem [{\citenamefont {Corkum}(1993)}]{Corkum_1993}%
  \BibitemOpen
  \bibfield  {author} {\bibinfo {author} {\bibfnamefont {P.~B.}\ \bibnamefont
  {Corkum}},\ }\href@noop {} {\bibfield  {journal} {\bibinfo  {journal} {Phys.
  Rev. Lett.}\ }\textbf {\bibinfo {volume} {71}},\ \bibinfo {pages} {1994}
  (\bibinfo {year} {1993})}\BibitemShut {NoStop}%
\bibitem [{\citenamefont {Stroke}(1966)}]{Stroke_1966}%
  \BibitemOpen
  \bibfield  {author} {\bibinfo {author} {\bibfnamefont {G.~W.}\ \bibnamefont
  {Stroke}},\ }\href@noop {} {\emph {\bibinfo {title} {An introduction to
  coherent optics and holography}}}\ (\bibinfo  {publisher} {Academic Press,
  New York},\ \bibinfo {year} {1966})\BibitemShut {NoStop}%
\bibitem [{\citenamefont {Hickstein}\ \emph {et~al.}(2012)\citenamefont
  {Hickstein}, \citenamefont {Ranitovic}, \citenamefont {Witte}, \citenamefont
  {Tong}, \citenamefont {Huismans}, \citenamefont {Arpin}, \citenamefont
  {Zhou}, \citenamefont {Keister}, \citenamefont {Hogle}, \citenamefont
  {Zhang}, \citenamefont {Ding}, \citenamefont {Johnsson}, \citenamefont
  {Toshima}, \citenamefont {Vrakking}, \citenamefont {Murnane},\ and\
  \citenamefont {Kapteyn}}]{Hickstein_2012}%
  \BibitemOpen
  \bibfield  {author} {\bibinfo {author} {\bibfnamefont {D.~D.}\ \bibnamefont
  {Hickstein}}, \bibinfo {author} {\bibfnamefont {P.}~\bibnamefont
  {Ranitovic}}, \bibinfo {author} {\bibfnamefont {S.}~\bibnamefont {Witte}},
  \bibinfo {author} {\bibfnamefont {X.-M.}\ \bibnamefont {Tong}}, \bibinfo
  {author} {\bibfnamefont {Y.}~\bibnamefont {Huismans}}, \bibinfo {author}
  {\bibfnamefont {P.}~\bibnamefont {Arpin}}, \bibinfo {author} {\bibfnamefont
  {X.}~\bibnamefont {Zhou}}, \bibinfo {author} {\bibfnamefont {K.~E.}\
  \bibnamefont {Keister}}, \bibinfo {author} {\bibfnamefont {C.~W.}\
  \bibnamefont {Hogle}}, \bibinfo {author} {\bibfnamefont {B.}~\bibnamefont
  {Zhang}}, \bibinfo {author} {\bibfnamefont {C.}~\bibnamefont {Ding}},
  \bibinfo {author} {\bibfnamefont {P.}~\bibnamefont {Johnsson}}, \bibinfo
  {author} {\bibfnamefont {N.}~\bibnamefont {Toshima}}, \bibinfo {author}
  {\bibfnamefont {M.~J.~J.}\ \bibnamefont {Vrakking}}, \bibinfo {author}
  {\bibfnamefont {M.~M.}\ \bibnamefont {Murnane}},\ and\ \bibinfo {author}
  {\bibfnamefont {H.~C.}\ \bibnamefont {Kapteyn}},\ }\href@noop {} {\bibfield
  {journal} {\bibinfo  {journal} {Phys. Rev. Lett.}\ }\textbf {\bibinfo
  {volume} {109}},\ \bibinfo {pages} {073004} (\bibinfo {year}
  {2012})}\BibitemShut {NoStop}%
\bibitem [{\citenamefont {Zhou}\ \emph {et~al.}(2016)\citenamefont {Zhou},
  \citenamefont {Tolstikhin},\ and\ \citenamefont {Morishita}}]{Zhou_2016}%
  \BibitemOpen
  \bibfield  {author} {\bibinfo {author} {\bibfnamefont {Y.}~\bibnamefont
  {Zhou}}, \bibinfo {author} {\bibfnamefont {O.~I.}\ \bibnamefont
  {Tolstikhin}},\ and\ \bibinfo {author} {\bibfnamefont {T.}~\bibnamefont
  {Morishita}},\ }\href@noop {} {\bibfield  {journal} {\bibinfo  {journal}
  {Phys. Rev. Lett.}\ }\textbf {\bibinfo {volume} {116}},\ \bibinfo {pages}
  {173001} (\bibinfo {year} {2016})}\BibitemShut {NoStop}%
\bibitem [{\citenamefont {He}\ \emph {et~al.}(2018)\citenamefont {He},
  \citenamefont {Li}, \citenamefont {Zhou}, \citenamefont {Li}, \citenamefont
  {Cao},\ and\ \citenamefont {Lu}}]{He_2018}%
  \BibitemOpen
  \bibfield  {author} {\bibinfo {author} {\bibfnamefont {M.}~\bibnamefont
  {He}}, \bibinfo {author} {\bibfnamefont {Y.}~\bibnamefont {Li}}, \bibinfo
  {author} {\bibfnamefont {Y.}~\bibnamefont {Zhou}}, \bibinfo {author}
  {\bibfnamefont {M.}~\bibnamefont {Li}}, \bibinfo {author} {\bibfnamefont
  {W.}~\bibnamefont {Cao}},\ and\ \bibinfo {author} {\bibfnamefont
  {P.}~\bibnamefont {Lu}},\ }\href@noop {} {\bibfield  {journal} {\bibinfo
  {journal} {Phys. Rev. Lett.}\ }\textbf {\bibinfo {volume} {120}},\ \bibinfo
  {pages} {133204} (\bibinfo {year} {2018})}\BibitemShut {NoStop}%
\bibitem [{\citenamefont {Bian}\ and\ \citenamefont
  {Bandrauk}(2012)}]{Bian_2012}%
  \BibitemOpen
  \bibfield  {author} {\bibinfo {author} {\bibfnamefont {X.-B.}\ \bibnamefont
  {Bian}}\ and\ \bibinfo {author} {\bibfnamefont {A.~D.}\ \bibnamefont
  {Bandrauk}},\ }\href@noop {} {\bibfield  {journal} {\bibinfo  {journal}
  {Phys. Rev. Lett.}\ }\textbf {\bibinfo {volume} {108}},\ \bibinfo {pages}
  {263003} (\bibinfo {year} {2012})}\BibitemShut {NoStop}%
\bibitem [{\citenamefont {Bian}\ and\ \citenamefont
  {Bandrauk}(2014)}]{Bian_2014}%
  \BibitemOpen
  \bibfield  {author} {\bibinfo {author} {\bibfnamefont {X.-B.}\ \bibnamefont
  {Bian}}\ and\ \bibinfo {author} {\bibfnamefont {A.~D.}\ \bibnamefont
  {Bandrauk}},\ }\href@noop {} {\bibfield  {journal} {\bibinfo  {journal}
  {Phys. Rev. A}\ }\textbf {\bibinfo {volume} {89}},\ \bibinfo {pages} {033423}
  (\bibinfo {year} {2014})}\BibitemShut {NoStop}%
\bibitem [{\citenamefont {Meckel}\ \emph {et~al.}(2014)\citenamefont {Meckel},
  \citenamefont {Staudte}, \citenamefont {Patchkovskii}, \citenamefont
  {Villeneuve}, \citenamefont {Corkum}, \citenamefont {D{\"{o}}rner},\ and\
  \citenamefont {Spanner}}]{Meckel_2014}%
  \BibitemOpen
  \bibfield  {author} {\bibinfo {author} {\bibfnamefont {M.}~\bibnamefont
  {Meckel}}, \bibinfo {author} {\bibfnamefont {A.}~\bibnamefont {Staudte}},
  \bibinfo {author} {\bibfnamefont {S.}~\bibnamefont {Patchkovskii}}, \bibinfo
  {author} {\bibfnamefont {D.~M.}\ \bibnamefont {Villeneuve}}, \bibinfo
  {author} {\bibfnamefont {P.~B.}\ \bibnamefont {Corkum}}, \bibinfo {author}
  {\bibfnamefont {R.}~\bibnamefont {D{\"{o}}rner}},\ and\ \bibinfo {author}
  {\bibfnamefont {M.}~\bibnamefont {Spanner}},\ }\href@noop {} {\bibfield
  {journal} {\bibinfo  {journal} {Nature Phys.}\ }\textbf {\bibinfo {volume}
  {10}},\ \bibinfo {pages} {594} (\bibinfo {year} {2014})}\BibitemShut
  {NoStop}%
\bibitem [{\citenamefont {Haertelt}\ \emph {et~al.}(2016)\citenamefont
  {Haertelt}, \citenamefont {Bian}, \citenamefont {Spanner}, \citenamefont
  {Staudte},\ and\ \citenamefont {Corkum}}]{Haertelt_2016}%
  \BibitemOpen
  \bibfield  {author} {\bibinfo {author} {\bibfnamefont {M.}~\bibnamefont
  {Haertelt}}, \bibinfo {author} {\bibfnamefont {X.-B.}\ \bibnamefont {Bian}},
  \bibinfo {author} {\bibfnamefont {M.}~\bibnamefont {Spanner}}, \bibinfo
  {author} {\bibfnamefont {A.}~\bibnamefont {Staudte}},\ and\ \bibinfo {author}
  {\bibfnamefont {P.~B.}\ \bibnamefont {Corkum}},\ }\href@noop {} {\bibfield
  {journal} {\bibinfo  {journal} {Phys. Rev. Lett.}\ }\textbf {\bibinfo
  {volume} {116}},\ \bibinfo {pages} {133001} (\bibinfo {year}
  {2016})}\BibitemShut {NoStop}%
\bibitem [{\citenamefont {Walt}\ \emph {et~al.}(2017)\citenamefont {Walt},
  \citenamefont {{Bhargava Ram}}, \citenamefont {Atala}, \citenamefont
  {Shvetsov-Shilovski}, \citenamefont {{Von Conta}}, \citenamefont
  {Baykusheva}, \citenamefont {Lein},\ and\ \citenamefont
  {W{\"{o}}rner}}]{Walt_2017}%
  \BibitemOpen
  \bibfield  {author} {\bibinfo {author} {\bibfnamefont {S.~G.}\ \bibnamefont
  {Walt}}, \bibinfo {author} {\bibfnamefont {N.}~\bibnamefont {{Bhargava
  Ram}}}, \bibinfo {author} {\bibfnamefont {M.}~\bibnamefont {Atala}}, \bibinfo
  {author} {\bibfnamefont {N.~I.}\ \bibnamefont {Shvetsov-Shilovski}}, \bibinfo
  {author} {\bibfnamefont {A.}~\bibnamefont {{Von Conta}}}, \bibinfo {author}
  {\bibfnamefont {D.}~\bibnamefont {Baykusheva}}, \bibinfo {author}
  {\bibfnamefont {M.}~\bibnamefont {Lein}},\ and\ \bibinfo {author}
  {\bibfnamefont {H.~J.}\ \bibnamefont {W{\"{o}}rner}},\ }\href@noop {}
  {\bibfield  {journal} {\bibinfo  {journal} {Nature Commun.}\ }\textbf
  {\bibinfo {volume} {8}},\ \bibinfo {pages} {15651 EP } (\bibinfo {year}
  {2017})}\BibitemShut {NoStop}%
\bibitem [{\citenamefont {Chelkowski}\ \emph {et~al.}(2015)\citenamefont
  {Chelkowski}, \citenamefont {Bandrauk},\ and\ \citenamefont
  {Corkum}}]{Chelkowski_2015}%
  \BibitemOpen
  \bibfield  {author} {\bibinfo {author} {\bibfnamefont {S.}~\bibnamefont
  {Chelkowski}}, \bibinfo {author} {\bibfnamefont {A.~D.}\ \bibnamefont
  {Bandrauk}},\ and\ \bibinfo {author} {\bibfnamefont {P.~B.}\ \bibnamefont
  {Corkum}},\ }\href@noop {} {\bibfield  {journal} {\bibinfo  {journal} {Phys.
  Rev. A}\ }\textbf {\bibinfo {volume} {92}},\ \bibinfo {pages} {051401}
  (\bibinfo {year} {2015})}\BibitemShut {NoStop}%
\bibitem [{\citenamefont {Ivanov}\ \emph {et~al.}(2016)\citenamefont {Ivanov},
  \citenamefont {Dubau},\ and\ \citenamefont {Kim}}]{Ivanov_2016}%
  \BibitemOpen
  \bibfield  {author} {\bibinfo {author} {\bibfnamefont {I.~A.}\ \bibnamefont
  {Ivanov}}, \bibinfo {author} {\bibfnamefont {J.}~\bibnamefont {Dubau}},\ and\
  \bibinfo {author} {\bibfnamefont {K.~T.}\ \bibnamefont {Kim}},\ }\href@noop
  {} {\bibfield  {journal} {\bibinfo  {journal} {Phys. Rev. A}\ }\textbf
  {\bibinfo {volume} {94}},\ \bibinfo {pages} {033405} (\bibinfo {year}
  {2016})}\BibitemShut {NoStop}%
\bibitem [{\citenamefont {Brennecke}\ and\ \citenamefont
  {Lein}(2018{\natexlab{a}})}]{Brennecke_2018}%
  \BibitemOpen
  \bibfield  {author} {\bibinfo {author} {\bibfnamefont {S.}~\bibnamefont
  {Brennecke}}\ and\ \bibinfo {author} {\bibfnamefont {M.}~\bibnamefont
  {Lein}},\ }\href@noop {} {\bibfield  {journal} {\bibinfo  {journal} {J. Phys.
  B}\ }\textbf {\bibinfo {volume} {51}},\ \bibinfo {pages} {094005} (\bibinfo
  {year} {2018}{\natexlab{a}})}\BibitemShut {NoStop}%
\bibitem [{\citenamefont {Brennecke}\ and\ \citenamefont
  {Lein}(2018{\natexlab{b}})}]{Brennecke_2018a}%
  \BibitemOpen
  \bibfield  {author} {\bibinfo {author} {\bibfnamefont {S.}~\bibnamefont
  {Brennecke}}\ and\ \bibinfo {author} {\bibfnamefont {M.}~\bibnamefont
  {Lein}},\ }\href@noop {} {\bibfield  {journal} {\bibinfo  {journal} {Phys.
  Rev. A}\ }\textbf {\bibinfo {volume} {98}},\ \bibinfo {pages} {063414}
  (\bibinfo {year} {2018}{\natexlab{b}})}\BibitemShut {NoStop}%
\bibitem [{\citenamefont {Brennecke}\ and\ \citenamefont
  {Lein}(2019)}]{Brennecke_2019}%
  \BibitemOpen
  \bibfield  {author} {\bibinfo {author} {\bibfnamefont {S.}~\bibnamefont
  {Brennecke}}\ and\ \bibinfo {author} {\bibfnamefont {M.}~\bibnamefont
  {Lein}},\ }\href@noop {} {\bibfield  {journal} {\bibinfo  {journal}
  {arxiv:1905.08143}\ } (\bibinfo {year} {2019})}\BibitemShut {NoStop}%
\bibitem [{\citenamefont {Smeenk}\ \emph
  {et~al.}(2011{\natexlab{a}})\citenamefont {Smeenk}, \citenamefont {Arissian},
  \citenamefont {Zhou}, \citenamefont {Mysyrowicz}, \citenamefont {Villeneuve},
  \citenamefont {Staudte},\ and\ \citenamefont {Corkum}}]{Smeenk_2011}%
  \BibitemOpen
  \bibfield  {author} {\bibinfo {author} {\bibfnamefont {C.~T.~L.}\
  \bibnamefont {Smeenk}}, \bibinfo {author} {\bibfnamefont {L.}~\bibnamefont
  {Arissian}}, \bibinfo {author} {\bibfnamefont {B.}~\bibnamefont {Zhou}},
  \bibinfo {author} {\bibfnamefont {A.}~\bibnamefont {Mysyrowicz}}, \bibinfo
  {author} {\bibfnamefont {D.~M.}\ \bibnamefont {Villeneuve}}, \bibinfo
  {author} {\bibfnamefont {A.}~\bibnamefont {Staudte}},\ and\ \bibinfo {author}
  {\bibfnamefont {P.~B.}\ \bibnamefont {Corkum}},\ }\href@noop {} {\bibfield
  {journal} {\bibinfo  {journal} {Phys. Rev. Lett.}\ }\textbf {\bibinfo
  {volume} {106}},\ \bibinfo {pages} {193002} (\bibinfo {year}
  {2011}{\natexlab{a}})}\BibitemShut {NoStop}%
\bibitem [{\citenamefont {Willenberg}\ \emph {et~al.}(2019)\citenamefont
  {Willenberg}, \citenamefont {Maurer}, \citenamefont {Mayer},\ and\
  \citenamefont {Keller}}]{Willenberg_2019}%
  \BibitemOpen
  \bibfield  {author} {\bibinfo {author} {\bibfnamefont {B.}~\bibnamefont
  {Willenberg}}, \bibinfo {author} {\bibfnamefont {J.}~\bibnamefont {Maurer}},
  \bibinfo {author} {\bibfnamefont {B.~W.}\ \bibnamefont {Mayer}},\ and\
  \bibinfo {author} {\bibfnamefont {U.}~\bibnamefont {Keller}},\ }\href@noop {}
  {\bibfield  {journal} {\bibinfo  {journal} {arxiv.org}\ ,\ \bibinfo {pages}
  {arXiv:1905.09546}} (\bibinfo {year} {2019})}\BibitemShut {NoStop}%
\bibitem [{\citenamefont {Titi}\ and\ \citenamefont {Drake}(2012)}]{Titi_2012}%
  \BibitemOpen
  \bibfield  {author} {\bibinfo {author} {\bibfnamefont {A.~S.}\ \bibnamefont
  {Titi}}\ and\ \bibinfo {author} {\bibfnamefont {G.~W.~F.}\ \bibnamefont
  {Drake}},\ }\href@noop {} {\bibfield  {journal} {\bibinfo  {journal} {Phys.
  Rev. A}\ }\textbf {\bibinfo {volume} {85}},\ \bibinfo {pages} {041404}
  (\bibinfo {year} {2012})}\BibitemShut {NoStop}%
\bibitem [{\citenamefont {Klaiber}\ \emph {et~al.}(2013)\citenamefont
  {Klaiber}, \citenamefont {Yakaboylu}, \citenamefont {Bauke}, \citenamefont
  {Hatsagortsyan},\ and\ \citenamefont {Keitel}}]{Klaiber_2013c}%
  \BibitemOpen
  \bibfield  {author} {\bibinfo {author} {\bibfnamefont {M.}~\bibnamefont
  {Klaiber}}, \bibinfo {author} {\bibfnamefont {E.}~\bibnamefont {Yakaboylu}},
  \bibinfo {author} {\bibfnamefont {H.}~\bibnamefont {Bauke}}, \bibinfo
  {author} {\bibfnamefont {K.~Z.}\ \bibnamefont {Hatsagortsyan}},\ and\
  \bibinfo {author} {\bibfnamefont {C.~H.}\ \bibnamefont {Keitel}},\
  }\href@noop {} {\bibfield  {journal} {\bibinfo  {journal} {Phys. Rev. Lett.}\
  }\textbf {\bibinfo {volume} {110}},\ \bibinfo {pages} {153004} (\bibinfo
  {year} {2013})}\BibitemShut {NoStop}%
\bibitem [{\citenamefont {Cricchio}\ \emph {et~al.}(2015)\citenamefont
  {Cricchio}, \citenamefont {Fiordilino},\ and\ \citenamefont
  {Hatsagortsyan}}]{Cricchio_2015}%
  \BibitemOpen
  \bibfield  {author} {\bibinfo {author} {\bibfnamefont {D.}~\bibnamefont
  {Cricchio}}, \bibinfo {author} {\bibfnamefont {E.}~\bibnamefont
  {Fiordilino}},\ and\ \bibinfo {author} {\bibfnamefont {K.~Z.}\ \bibnamefont
  {Hatsagortsyan}},\ }\href@noop {} {\bibfield  {journal} {\bibinfo  {journal}
  {Phys. Rev. A}\ }\textbf {\bibinfo {volume} {92}},\ \bibinfo {pages} {023408}
  (\bibinfo {year} {2015})}\BibitemShut {NoStop}%
\bibitem [{\citenamefont {Chelkowski}\ \emph {et~al.}(2017)\citenamefont
  {Chelkowski}, \citenamefont {Bandrauk},\ and\ \citenamefont
  {Corkum}}]{Chelkowski_2017}%
  \BibitemOpen
  \bibfield  {author} {\bibinfo {author} {\bibfnamefont {S.}~\bibnamefont
  {Chelkowski}}, \bibinfo {author} {\bibfnamefont {A.~D.}\ \bibnamefont
  {Bandrauk}},\ and\ \bibinfo {author} {\bibfnamefont {P.~B.}\ \bibnamefont
  {Corkum}},\ }\href {https://doi.org/10.1103/PhysRevA.95.053402} {\bibfield
  {journal} {\bibinfo  {journal} {Phys. Rev. A}\ }\textbf {\bibinfo {volume}
  {95}},\ \bibinfo {pages} {053402} (\bibinfo {year} {2017})}\BibitemShut
  {NoStop}%
\bibitem [{\citenamefont {He}\ \emph {et~al.}(2017)\citenamefont {He},
  \citenamefont {Lao},\ and\ \citenamefont {He}}]{He_2017}%
  \BibitemOpen
  \bibfield  {author} {\bibinfo {author} {\bibfnamefont {P.-L.}\ \bibnamefont
  {He}}, \bibinfo {author} {\bibfnamefont {D.}~\bibnamefont {Lao}},\ and\
  \bibinfo {author} {\bibfnamefont {F.}~\bibnamefont {He}},\ }\href@noop {}
  {\bibfield  {journal} {\bibinfo  {journal} {Phys. Rev. Lett.}\ }\textbf
  {\bibinfo {volume} {118}},\ \bibinfo {pages} {163203} (\bibinfo {year}
  {2017})}\BibitemShut {NoStop}%
\bibitem [{\citenamefont {Chelkowski}\ and\ \citenamefont
  {Bandrauk}(2018)}]{Chelkowski_2018}%
  \BibitemOpen
  \bibfield  {author} {\bibinfo {author} {\bibfnamefont {S.}~\bibnamefont
  {Chelkowski}}\ and\ \bibinfo {author} {\bibfnamefont {A.~D.}\ \bibnamefont
  {Bandrauk}},\ }\href@noop {} {\bibfield  {journal} {\bibinfo  {journal}
  {Phys. Rev. A}\ }\textbf {\bibinfo {volume} {97}},\ \bibinfo {pages} {053401}
  (\bibinfo {year} {2018})}\BibitemShut {NoStop}%
\bibitem [{\citenamefont {Dammasch}\ \emph {et~al.}(2001)\citenamefont
  {Dammasch}, \citenamefont {D\"orr}, \citenamefont {Eichmann}, \citenamefont
  {Lenz},\ and\ \citenamefont {Sandner}}]{Dammasch_2001}%
  \BibitemOpen
  \bibfield  {author} {\bibinfo {author} {\bibfnamefont {M.}~\bibnamefont
  {Dammasch}}, \bibinfo {author} {\bibfnamefont {M.}~\bibnamefont {D\"orr}},
  \bibinfo {author} {\bibfnamefont {U.}~\bibnamefont {Eichmann}}, \bibinfo
  {author} {\bibfnamefont {E.}~\bibnamefont {Lenz}},\ and\ \bibinfo {author}
  {\bibfnamefont {W.}~\bibnamefont {Sandner}},\ }\href@noop {} {\bibfield
  {journal} {\bibinfo  {journal} {Phys. Rev. A}\ }\textbf {\bibinfo {volume}
  {64}},\ \bibinfo {pages} {061402} (\bibinfo {year} {2001})}\BibitemShut
  {NoStop}%
\bibitem [{\citenamefont {Walser}\ \emph {et~al.}(2000)\citenamefont {Walser},
  \citenamefont {Keitel}, \citenamefont {Scrinzi},\ and\ \citenamefont
  {Brabec}}]{Walser_2000a}%
  \BibitemOpen
  \bibfield  {author} {\bibinfo {author} {\bibfnamefont {M.~W.}\ \bibnamefont
  {Walser}}, \bibinfo {author} {\bibfnamefont {C.~H.}\ \bibnamefont {Keitel}},
  \bibinfo {author} {\bibfnamefont {A.}~\bibnamefont {Scrinzi}},\ and\ \bibinfo
  {author} {\bibfnamefont {T.}~\bibnamefont {Brabec}},\ }\href@noop {}
  {\bibfield  {journal} {\bibinfo  {journal} {Phys. Rev. Lett.}\ }\textbf
  {\bibinfo {volume} {85}},\ \bibinfo {pages} {5082} (\bibinfo {year}
  {2000})}\BibitemShut {NoStop}%
\bibitem [{\citenamefont {Milo\v{s}evi\'{c}}\ \emph {et~al.}(2000)\citenamefont
  {Milo\v{s}evi\'{c}}, \citenamefont {Hu},\ and\ \citenamefont
  {Becker}}]{Milosevic_2000}%
  \BibitemOpen
  \bibfield  {author} {\bibinfo {author} {\bibfnamefont {D.~B.}\ \bibnamefont
  {Milo\v{s}evi\'{c}}}, \bibinfo {author} {\bibfnamefont {S.~X.}\ \bibnamefont
  {Hu}},\ and\ \bibinfo {author} {\bibfnamefont {W.}~\bibnamefont {Becker}},\
  }\href@noop {} {\bibfield  {journal} {\bibinfo  {journal} {Phys. Rev. A}\
  }\textbf {\bibinfo {volume} {63}},\ \bibinfo {pages} {011403(R)} (\bibinfo
  {year} {2000})}\BibitemShut {NoStop}%
\bibitem [{\citenamefont {Chiril\u{a}}\ \emph {et~al.}(2002)\citenamefont
  {Chiril\u{a}}, \citenamefont {Kylstra}, \citenamefont {Potvliege},\ and\
  \citenamefont {Joachain}}]{Chirila_2002}%
  \BibitemOpen
  \bibfield  {author} {\bibinfo {author} {\bibfnamefont {C.~C.}\ \bibnamefont
  {Chiril\u{a}}}, \bibinfo {author} {\bibfnamefont {N.~J.}\ \bibnamefont
  {Kylstra}}, \bibinfo {author} {\bibfnamefont {R.~M.}\ \bibnamefont
  {Potvliege}},\ and\ \bibinfo {author} {\bibfnamefont {C.~J.}\ \bibnamefont
  {Joachain}},\ }\href@noop {} {\bibfield  {journal} {\bibinfo  {journal}
  {Phys. Rev. A}\ }\textbf {\bibinfo {volume} {66}},\ \bibinfo {pages} {063411}
  (\bibinfo {year} {2002})}\BibitemShut {NoStop}%
\bibitem [{\citenamefont {Klaiber}\ \emph {et~al.}(2005)\citenamefont
  {Klaiber}, \citenamefont {Hatsagortsyan},\ and\ \citenamefont
  {Keitel}}]{Klaiber_2005}%
  \BibitemOpen
  \bibfield  {author} {\bibinfo {author} {\bibfnamefont {M.}~\bibnamefont
  {Klaiber}}, \bibinfo {author} {\bibfnamefont {K.~Z.}\ \bibnamefont
  {Hatsagortsyan}},\ and\ \bibinfo {author} {\bibfnamefont {C.~H.}\
  \bibnamefont {Keitel}},\ }\href@noop {} {\bibfield  {journal} {\bibinfo
  {journal} {Phys. Rev. A}\ }\textbf {\bibinfo {volume} {71}},\ \bibinfo
  {pages} {033408} (\bibinfo {year} {2005})}\BibitemShut {NoStop}%
\bibitem [{\citenamefont {Kohler}\ \emph {et~al.}(2012)\citenamefont {Kohler},
  \citenamefont {Pfeifer}, \citenamefont {Hatsagortsyan},\ and\ \citenamefont
  {Keitel}}]{Kohler_2012b}%
  \BibitemOpen
  \bibfield  {author} {\bibinfo {author} {\bibfnamefont {M.~C.}\ \bibnamefont
  {Kohler}}, \bibinfo {author} {\bibfnamefont {T.}~\bibnamefont {Pfeifer}},
  \bibinfo {author} {\bibfnamefont {K.~Z.}\ \bibnamefont {Hatsagortsyan}},\
  and\ \bibinfo {author} {\bibfnamefont {C.~H.}\ \bibnamefont {Keitel}},\
  }\href@noop {} {\bibfield  {journal} {\bibinfo  {journal} {Adv. At. Mol.
  Phys.}\ }\textbf {\bibinfo {volume} {61}},\ \bibinfo {pages} {159} (\bibinfo
  {year} {2012})}\BibitemShut {NoStop}%
\bibitem [{\citenamefont {Ludwig}\ \emph {et~al.}(2014)\citenamefont {Ludwig},
  \citenamefont {Maurer}, \citenamefont {Mayer}, \citenamefont {Phillips},
  \citenamefont {Gallmann},\ and\ \citenamefont {Keller}}]{Ludwig_2014}%
  \BibitemOpen
  \bibfield  {author} {\bibinfo {author} {\bibfnamefont {A.}~\bibnamefont
  {Ludwig}}, \bibinfo {author} {\bibfnamefont {J.}~\bibnamefont {Maurer}},
  \bibinfo {author} {\bibfnamefont {B.~W.}\ \bibnamefont {Mayer}}, \bibinfo
  {author} {\bibfnamefont {C.~R.}\ \bibnamefont {Phillips}}, \bibinfo {author}
  {\bibfnamefont {L.}~\bibnamefont {Gallmann}},\ and\ \bibinfo {author}
  {\bibfnamefont {U.}~\bibnamefont {Keller}},\ }\href@noop {} {\bibfield
  {journal} {\bibinfo  {journal} {Phys. Rev. Lett.}\ }\textbf {\bibinfo
  {volume} {113}},\ \bibinfo {pages} {243001} (\bibinfo {year}
  {2014})}\BibitemShut {NoStop}%
\bibitem [{\citenamefont {Brabec}\ \emph {et~al.}(1996)\citenamefont {Brabec},
  \citenamefont {Ivanov},\ and\ \citenamefont {Corkum}}]{Brabec_1996}%
  \BibitemOpen
  \bibfield  {author} {\bibinfo {author} {\bibfnamefont {T.}~\bibnamefont
  {Brabec}}, \bibinfo {author} {\bibfnamefont {M.~Y.}\ \bibnamefont {Ivanov}},\
  and\ \bibinfo {author} {\bibfnamefont {P.~B.}\ \bibnamefont {Corkum}},\
  }\href@noop {} {\bibfield  {journal} {\bibinfo  {journal} {Phys. Rev. A}\
  }\textbf {\bibinfo {volume} {54}},\ \bibinfo {pages} {R2551} (\bibinfo {year}
  {1996})}\BibitemShut {NoStop}%
\bibitem [{\citenamefont {F\o{}rre}\ \emph {et~al.}(2006)\citenamefont
  {F\o{}rre}, \citenamefont {Hansen}, \citenamefont {Kocbach}, \citenamefont
  {Selst\o{}},\ and\ \citenamefont {Madsen}}]{Foerre_2006}%
  \BibitemOpen
  \bibfield  {author} {\bibinfo {author} {\bibfnamefont {M.}~\bibnamefont
  {F\o{}rre}}, \bibinfo {author} {\bibfnamefont {J.~P.}\ \bibnamefont
  {Hansen}}, \bibinfo {author} {\bibfnamefont {L.}~\bibnamefont {Kocbach}},
  \bibinfo {author} {\bibfnamefont {S.}~\bibnamefont {Selst\o{}}},\ and\
  \bibinfo {author} {\bibfnamefont {L.~B.}\ \bibnamefont {Madsen}},\
  }\href@noop {} {\bibfield  {journal} {\bibinfo  {journal} {Phys. Rev. Lett.}\
  }\textbf {\bibinfo {volume} {97}},\ \bibinfo {pages} {043601} (\bibinfo
  {year} {2006})}\BibitemShut {NoStop}%
\bibitem [{\citenamefont {Keil}\ and\ \citenamefont {Bauer}(2017)}]{Keil_2017}%
  \BibitemOpen
  \bibfield  {author} {\bibinfo {author} {\bibfnamefont {T.}~\bibnamefont
  {Keil}}\ and\ \bibinfo {author} {\bibfnamefont {D.}~\bibnamefont {Bauer}},\
  }\href@noop {} {\bibfield  {journal} {\bibinfo  {journal} {J. Phys. B}\
  }\textbf {\bibinfo {volume} {50}},\ \bibinfo {pages} {194002} (\bibinfo
  {year} {2017})}\BibitemShut {NoStop}%
\bibitem [{\citenamefont {Tao}\ \emph {et~al.}(2017)\citenamefont {Tao},
  \citenamefont {Xia}, \citenamefont {Cai}, \citenamefont {Fu},\ and\
  \citenamefont {Liu}}]{Tao_2017}%
  \BibitemOpen
  \bibfield  {author} {\bibinfo {author} {\bibfnamefont {J.~F.}\ \bibnamefont
  {Tao}}, \bibinfo {author} {\bibfnamefont {Q.~Z.}\ \bibnamefont {Xia}},
  \bibinfo {author} {\bibfnamefont {J.}~\bibnamefont {Cai}}, \bibinfo {author}
  {\bibfnamefont {L.~B.}\ \bibnamefont {Fu}},\ and\ \bibinfo {author}
  {\bibfnamefont {J.}~\bibnamefont {Liu}},\ }\href@noop {} {\bibfield
  {journal} {\bibinfo  {journal} {Phys. Rev. A}\ }\textbf {\bibinfo {volume}
  {95}},\ \bibinfo {pages} {011402} (\bibinfo {year} {2017})}\BibitemShut
  {NoStop}%
\bibitem [{\citenamefont {Dan\ifmmode~\check{e}\else \v{e}\fi{}k}\ \emph
  {et~al.}(2018{\natexlab{a}})\citenamefont {Dan\ifmmode~\check{e}\else
  \v{e}\fi{}k}, \citenamefont {Hatsagortsyan},\ and\ \citenamefont
  {Keitel}}]{Danek_2018}%
  \BibitemOpen
  \bibfield  {author} {\bibinfo {author} {\bibfnamefont {J.}~\bibnamefont
  {Dan\ifmmode~\check{e}\else \v{e}\fi{}k}}, \bibinfo {author} {\bibfnamefont
  {K.~Z.}\ \bibnamefont {Hatsagortsyan}},\ and\ \bibinfo {author}
  {\bibfnamefont {C.~H.}\ \bibnamefont {Keitel}},\ }\href@noop {} {\bibfield
  {journal} {\bibinfo  {journal} {Phys. Rev. A}\ }\textbf {\bibinfo {volume}
  {97}},\ \bibinfo {pages} {063409} (\bibinfo {year}
  {2018}{\natexlab{a}})}\BibitemShut {NoStop}%
\bibitem [{\citenamefont {Maurer}\ \emph {et~al.}(2018)\citenamefont {Maurer},
  \citenamefont {Willenberg}, \citenamefont {Dan\ifmmode~\check{e}\else
  \v{e}\fi{}k}, \citenamefont {Mayer}, \citenamefont {Phillips}, \citenamefont
  {Gallmann}, \citenamefont {Klaiber}, \citenamefont {Hatsagortsyan},
  \citenamefont {Keitel},\ and\ \citenamefont {Keller}}]{Maurer_2018}%
  \BibitemOpen
  \bibfield  {author} {\bibinfo {author} {\bibfnamefont {J.}~\bibnamefont
  {Maurer}}, \bibinfo {author} {\bibfnamefont {B.}~\bibnamefont {Willenberg}},
  \bibinfo {author} {\bibfnamefont {J.}~\bibnamefont
  {Dan\ifmmode~\check{e}\else \v{e}\fi{}k}}, \bibinfo {author} {\bibfnamefont
  {B.~W.}\ \bibnamefont {Mayer}}, \bibinfo {author} {\bibfnamefont {C.~R.}\
  \bibnamefont {Phillips}}, \bibinfo {author} {\bibfnamefont {L.}~\bibnamefont
  {Gallmann}}, \bibinfo {author} {\bibfnamefont {M.}~\bibnamefont {Klaiber}},
  \bibinfo {author} {\bibfnamefont {K.~Z.}\ \bibnamefont {Hatsagortsyan}},
  \bibinfo {author} {\bibfnamefont {C.~H.}\ \bibnamefont {Keitel}},\ and\
  \bibinfo {author} {\bibfnamefont {U.}~\bibnamefont {Keller}},\ }\href@noop {}
  {\bibfield  {journal} {\bibinfo  {journal} {Phys. Rev. A}\ }\textbf {\bibinfo
  {volume} {97}},\ \bibinfo {pages} {013404} (\bibinfo {year}
  {2018})}\BibitemShut {NoStop}%
\bibitem [{\citenamefont {Dan\ifmmode~\check{e}\else \v{e}\fi{}k}\ \emph
  {et~al.}(2018{\natexlab{b}})\citenamefont {Dan\ifmmode~\check{e}\else
  \v{e}\fi{}k}, \citenamefont {Klaiber}, \citenamefont {Hatsagortsyan},
  \citenamefont {Keitel}, \citenamefont {Willenberg}, \citenamefont {Maurer},
  \citenamefont {Mayer}, \citenamefont {Phillips}, \citenamefont {Gallmann},\
  and\ \citenamefont {Keller}}]{Danek_2018b}%
  \BibitemOpen
  \bibfield  {author} {\bibinfo {author} {\bibfnamefont {J.}~\bibnamefont
  {Dan\ifmmode~\check{e}\else \v{e}\fi{}k}}, \bibinfo {author} {\bibfnamefont
  {M.}~\bibnamefont {Klaiber}}, \bibinfo {author} {\bibfnamefont {K.~Z.}\
  \bibnamefont {Hatsagortsyan}}, \bibinfo {author} {\bibfnamefont {C.~H.}\
  \bibnamefont {Keitel}}, \bibinfo {author} {\bibfnamefont {B.}~\bibnamefont
  {Willenberg}}, \bibinfo {author} {\bibfnamefont {J.}~\bibnamefont {Maurer}},
  \bibinfo {author} {\bibfnamefont {B.~W.}\ \bibnamefont {Mayer}}, \bibinfo
  {author} {\bibfnamefont {C.~R.}\ \bibnamefont {Phillips}}, \bibinfo {author}
  {\bibfnamefont {L.}~\bibnamefont {Gallmann}},\ and\ \bibinfo {author}
  {\bibfnamefont {U.}~\bibnamefont {Keller}},\ }\href@noop {} {\bibfield
  {journal} {\bibinfo  {journal} {J. Phys. B}\ }\textbf {\bibinfo {volume}
  {51}},\ \bibinfo {pages} {114001} (\bibinfo {year}
  {2018}{\natexlab{b}})}\BibitemShut {NoStop}%
\bibitem [{\citenamefont {Palaniyappan}\ \emph {et~al.}(2006)\citenamefont
  {Palaniyappan}, \citenamefont {Ghebregziabher}, \citenamefont {DiChiara},
  \citenamefont {MacDonald},\ and\ \citenamefont
  {Walker}}]{Palaniyappan_2006a}%
  \BibitemOpen
  \bibfield  {author} {\bibinfo {author} {\bibfnamefont {S.}~\bibnamefont
  {Palaniyappan}}, \bibinfo {author} {\bibfnamefont {I.}~\bibnamefont
  {Ghebregziabher}}, \bibinfo {author} {\bibfnamefont {A.~D.}\ \bibnamefont
  {DiChiara}}, \bibinfo {author} {\bibfnamefont {J.}~\bibnamefont
  {MacDonald}},\ and\ \bibinfo {author} {\bibfnamefont {B.~C.}\ \bibnamefont
  {Walker}},\ }\href@noop {} {\bibfield  {journal} {\bibinfo  {journal} {Phys.
  Rev. A}\ }\textbf {\bibinfo {volume} {74}},\ \bibinfo {pages} {033403}
  (\bibinfo {year} {2006})}\BibitemShut {NoStop}%
\bibitem [{\citenamefont {Reiss}(2008)}]{Reiss_2008}%
  \BibitemOpen
  \bibfield  {author} {\bibinfo {author} {\bibfnamefont {H.~R.}\ \bibnamefont
  {Reiss}},\ }\href@noop {} {\bibfield  {journal} {\bibinfo  {journal} {Phys.
  Rev. Lett.}\ }\textbf {\bibinfo {volume} {101}},\ \bibinfo {pages} {043002}
  (\bibinfo {year} {2008})}\BibitemShut {NoStop}%
\bibitem [{\citenamefont {Klaiber}\ \emph {et~al.}(2017)\citenamefont
  {Klaiber}, \citenamefont {Hatsagortsyan}, \citenamefont {Wu}, \citenamefont
  {Luo}, \citenamefont {Grugan},\ and\ \citenamefont {Walker}}]{Klaiber_2017}%
  \BibitemOpen
  \bibfield  {author} {\bibinfo {author} {\bibfnamefont {M.}~\bibnamefont
  {Klaiber}}, \bibinfo {author} {\bibfnamefont {K.~Z.}\ \bibnamefont
  {Hatsagortsyan}}, \bibinfo {author} {\bibfnamefont {J.}~\bibnamefont {Wu}},
  \bibinfo {author} {\bibfnamefont {S.~S.}\ \bibnamefont {Luo}}, \bibinfo
  {author} {\bibfnamefont {P.}~\bibnamefont {Grugan}},\ and\ \bibinfo {author}
  {\bibfnamefont {B.~C.}\ \bibnamefont {Walker}},\ }\href@noop {} {\bibfield
  {journal} {\bibinfo  {journal} {Phys. Rev. Lett.}\ }\textbf {\bibinfo
  {volume} {118}},\ \bibinfo {pages} {093001} (\bibinfo {year}
  {2017})}\BibitemShut {NoStop}%
\bibitem [{\citenamefont {McNaught}\ \emph {et~al.}(1997)\citenamefont
  {McNaught}, \citenamefont {Knauer},\ and\ \citenamefont
  {Meyerhofer}}]{McNaught_1997}%
  \BibitemOpen
  \bibfield  {author} {\bibinfo {author} {\bibfnamefont {S.~J.}\ \bibnamefont
  {McNaught}}, \bibinfo {author} {\bibfnamefont {J.~P.}\ \bibnamefont
  {Knauer}},\ and\ \bibinfo {author} {\bibfnamefont {D.~D.}\ \bibnamefont
  {Meyerhofer}},\ }\href@noop {} {\bibfield  {journal} {\bibinfo  {journal}
  {Phys. Rev. Lett.}\ }\textbf {\bibinfo {volume} {78}},\ \bibinfo {pages}
  {626} (\bibinfo {year} {1997})}\BibitemShut {NoStop}%
\bibitem [{\citenamefont {Moore}\ \emph {et~al.}(1999)\citenamefont {Moore},
  \citenamefont {Ting}, \citenamefont {McNaught}, \citenamefont {Qiu},
  \citenamefont {Burris},\ and\ \citenamefont {Sprangle}}]{Moore_1999}%
  \BibitemOpen
  \bibfield  {author} {\bibinfo {author} {\bibfnamefont {C.~I.}\ \bibnamefont
  {Moore}}, \bibinfo {author} {\bibfnamefont {A.}~\bibnamefont {Ting}},
  \bibinfo {author} {\bibfnamefont {S.~J.}\ \bibnamefont {McNaught}}, \bibinfo
  {author} {\bibfnamefont {J.}~\bibnamefont {Qiu}}, \bibinfo {author}
  {\bibfnamefont {H.~R.}\ \bibnamefont {Burris}},\ and\ \bibinfo {author}
  {\bibfnamefont {P.}~\bibnamefont {Sprangle}},\ }\href@noop {} {\bibfield
  {journal} {\bibinfo  {journal} {Phys. Rev. Lett.}\ }\textbf {\bibinfo
  {volume} {82}},\ \bibinfo {pages} {1688} (\bibinfo {year}
  {1999})}\BibitemShut {NoStop}%
\bibitem [{\citenamefont {Chowdhury}\ and\ \citenamefont
  {Walker}(2003)}]{Chowdhury2003}%
  \BibitemOpen
  \bibfield  {author} {\bibinfo {author} {\bibfnamefont {E.~A.}\ \bibnamefont
  {Chowdhury}}\ and\ \bibinfo {author} {\bibfnamefont {B.~C.}\ \bibnamefont
  {Walker}},\ }\href@noop {} {\bibfield  {journal} {\bibinfo  {journal} {J.
  Opt. Soc. Am. B}\ }\textbf {\bibinfo {volume} {20}},\ \bibinfo {pages} {109}
  (\bibinfo {year} {2003})}\BibitemShut {NoStop}%
\bibitem [{\citenamefont {Gubbini}\ \emph {et~al.}(2005)\citenamefont
  {Gubbini}, \citenamefont {Eichmann}, \citenamefont {Kalashnikov},\ and\
  \citenamefont {Sandner}}]{Gubbini_2005}%
  \BibitemOpen
  \bibfield  {author} {\bibinfo {author} {\bibfnamefont {E.}~\bibnamefont
  {Gubbini}}, \bibinfo {author} {\bibfnamefont {U.}~\bibnamefont {Eichmann}},
  \bibinfo {author} {\bibfnamefont {M.}~\bibnamefont {Kalashnikov}},\ and\
  \bibinfo {author} {\bibfnamefont {W.}~\bibnamefont {Sandner}},\ }\href@noop
  {} {\bibfield  {journal} {\bibinfo  {journal} {Phys. Rev. Lett.}\ }\textbf
  {\bibinfo {volume} {94}},\ \bibinfo {pages} {053602} (\bibinfo {year}
  {2005})}\BibitemShut {NoStop}%
\bibitem [{\citenamefont {Palaniyappan}\ \emph {et~al.}(2005)\citenamefont
  {Palaniyappan}, \citenamefont {DiChiara}, \citenamefont {Chowdhury},
  \citenamefont {Falkowski}, \citenamefont {Ongadi}, \citenamefont {Huskins},\
  and\ \citenamefont {Walker}}]{Palaniyappan2005}%
  \BibitemOpen
  \bibfield  {author} {\bibinfo {author} {\bibfnamefont {S.}~\bibnamefont
  {Palaniyappan}}, \bibinfo {author} {\bibfnamefont {A.}~\bibnamefont
  {DiChiara}}, \bibinfo {author} {\bibfnamefont {E.}~\bibnamefont {Chowdhury}},
  \bibinfo {author} {\bibfnamefont {A.}~\bibnamefont {Falkowski}}, \bibinfo
  {author} {\bibfnamefont {G.}~\bibnamefont {Ongadi}}, \bibinfo {author}
  {\bibfnamefont {E.~L.}\ \bibnamefont {Huskins}},\ and\ \bibinfo {author}
  {\bibfnamefont {B.~C.}\ \bibnamefont {Walker}},\ }\href@noop {} {\bibfield
  {journal} {\bibinfo  {journal} {Phys. Rev. Lett.}\ }\textbf {\bibinfo
  {volume} {94}},\ \bibinfo {pages} {243003} (\bibinfo {year}
  {2005})}\BibitemShut {NoStop}%
\bibitem [{\citenamefont {DiChiara}\ \emph {et~al.}(2008)\citenamefont
  {DiChiara}, \citenamefont {Ghebregziabher}, \citenamefont {Sauer},
  \citenamefont {Waesche}, \citenamefont {Palaniyappan}, \citenamefont {Wen},\
  and\ \citenamefont {Walker}}]{DiChiara_2008}%
  \BibitemOpen
  \bibfield  {author} {\bibinfo {author} {\bibfnamefont {A.~D.}\ \bibnamefont
  {DiChiara}}, \bibinfo {author} {\bibfnamefont {I.}~\bibnamefont
  {Ghebregziabher}}, \bibinfo {author} {\bibfnamefont {R.}~\bibnamefont
  {Sauer}}, \bibinfo {author} {\bibfnamefont {J.}~\bibnamefont {Waesche}},
  \bibinfo {author} {\bibfnamefont {S.}~\bibnamefont {Palaniyappan}}, \bibinfo
  {author} {\bibfnamefont {B.~L.}\ \bibnamefont {Wen}},\ and\ \bibinfo {author}
  {\bibfnamefont {B.~C.}\ \bibnamefont {Walker}},\ }\href@noop {} {\bibfield
  {journal} {\bibinfo  {journal} {Phys. Rev. Lett.}\ }\textbf {\bibinfo
  {volume} {101}},\ \bibinfo {pages} {173002} (\bibinfo {year}
  {2008})}\BibitemShut {NoStop}%
\bibitem [{\citenamefont {Palaniyappan}\ \emph {et~al.}(2008)\citenamefont
  {Palaniyappan}, \citenamefont {Mitchell}, \citenamefont {Sauer},
  \citenamefont {Ghebregziabher}, \citenamefont {White}, \citenamefont
  {Decamp},\ and\ \citenamefont {Walker}}]{Palaniyappan_2008}%
  \BibitemOpen
  \bibfield  {author} {\bibinfo {author} {\bibfnamefont {S.}~\bibnamefont
  {Palaniyappan}}, \bibinfo {author} {\bibfnamefont {R.}~\bibnamefont
  {Mitchell}}, \bibinfo {author} {\bibfnamefont {R.}~\bibnamefont {Sauer}},
  \bibinfo {author} {\bibfnamefont {I.}~\bibnamefont {Ghebregziabher}},
  \bibinfo {author} {\bibfnamefont {S.~L.}\ \bibnamefont {White}}, \bibinfo
  {author} {\bibfnamefont {M.~F.}\ \bibnamefont {Decamp}},\ and\ \bibinfo
  {author} {\bibfnamefont {B.~C.}\ \bibnamefont {Walker}},\ }\href@noop {}
  {\bibfield  {journal} {\bibinfo  {journal} {Phys. Rev. Lett.}\ }\textbf
  {\bibinfo {volume} {100}},\ \bibinfo {pages} {183001} (\bibinfo {year}
  {2008})}\BibitemShut {NoStop}%
\bibitem [{\citenamefont {Ekanayake}\ \emph {et~al.}(2013)\citenamefont
  {Ekanayake}, \citenamefont {Luo}, \citenamefont {Grugan}, \citenamefont
  {Crosby}, \citenamefont {Camilo}, \citenamefont {McCowan}, \citenamefont
  {Scalzi}, \citenamefont {Tramontozzi}, \citenamefont {Howard}, \citenamefont
  {Wells}, \citenamefont {Mancuso}, \citenamefont {Stanev}, \citenamefont
  {Decamp},\ and\ \citenamefont {Walker}}]{Ekanayake_2013}%
  \BibitemOpen
  \bibfield  {author} {\bibinfo {author} {\bibfnamefont {N.}~\bibnamefont
  {Ekanayake}}, \bibinfo {author} {\bibfnamefont {S.}~\bibnamefont {Luo}},
  \bibinfo {author} {\bibfnamefont {P.~D.}\ \bibnamefont {Grugan}}, \bibinfo
  {author} {\bibfnamefont {W.~B.}\ \bibnamefont {Crosby}}, \bibinfo {author}
  {\bibfnamefont {A.~D.}\ \bibnamefont {Camilo}}, \bibinfo {author}
  {\bibfnamefont {C.~V.}\ \bibnamefont {McCowan}}, \bibinfo {author}
  {\bibfnamefont {R.}~\bibnamefont {Scalzi}}, \bibinfo {author} {\bibfnamefont
  {A.}~\bibnamefont {Tramontozzi}}, \bibinfo {author} {\bibfnamefont {L.~E.}\
  \bibnamefont {Howard}}, \bibinfo {author} {\bibfnamefont {S.~J.}\
  \bibnamefont {Wells}}, \bibinfo {author} {\bibfnamefont {C.}~\bibnamefont
  {Mancuso}}, \bibinfo {author} {\bibfnamefont {T.}~\bibnamefont {Stanev}},
  \bibinfo {author} {\bibfnamefont {M.~F.}\ \bibnamefont {Decamp}},\ and\
  \bibinfo {author} {\bibfnamefont {B.~C.}\ \bibnamefont {Walker}},\
  }\href@noop {} {\bibfield  {journal} {\bibinfo  {journal} {Phys. Rev. Lett.}\
  }\textbf {\bibinfo {volume} {110}},\ \bibinfo {pages} {203003} (\bibinfo
  {year} {2013})}\BibitemShut {NoStop}%
\bibitem [{\citenamefont {Mayer}\ \emph {et~al.}(2014)\citenamefont {Mayer},
  \citenamefont {Phillips}, \citenamefont {Gallmann},\ and\ \citenamefont
  {Keller}}]{Mayer_2014OSA}%
  \BibitemOpen
  \bibfield  {author} {\bibinfo {author} {\bibfnamefont {B.~W.}\ \bibnamefont
  {Mayer}}, \bibinfo {author} {\bibfnamefont {C.~R.}\ \bibnamefont {Phillips}},
  \bibinfo {author} {\bibfnamefont {L.}~\bibnamefont {Gallmann}},\ and\
  \bibinfo {author} {\bibfnamefont {U.}~\bibnamefont {Keller}},\ }\href
  {https://doi.org/10.1364/OE.22.020798} {\bibfield  {journal} {\bibinfo
  {journal} {Opt. Expr.}\ }\textbf {\bibinfo {volume} {22}},\ \bibinfo {pages}
  {20798} (\bibinfo {year} {2014})}\BibitemShut {NoStop}%
\bibitem [{\citenamefont {Mayer}\ \emph {et~al.}(2013)\citenamefont {Mayer},
  \citenamefont {Phillips}, \citenamefont {Gallmann}, \citenamefont {Fejer},\
  and\ \citenamefont {Keller}}]{Mayer_2013}%
  \BibitemOpen
  \bibfield  {author} {\bibinfo {author} {\bibfnamefont {B.~W.}\ \bibnamefont
  {Mayer}}, \bibinfo {author} {\bibfnamefont {C.~R.}\ \bibnamefont {Phillips}},
  \bibinfo {author} {\bibfnamefont {L.}~\bibnamefont {Gallmann}}, \bibinfo
  {author} {\bibfnamefont {M.~M.}\ \bibnamefont {Fejer}},\ and\ \bibinfo
  {author} {\bibfnamefont {U.}~\bibnamefont {Keller}},\ }\href
  {https://doi.org/10.1364/OL.38.004265} {\bibfield  {journal} {\bibinfo
  {journal} {Opt. Lett.}\ }\textbf {\bibinfo {volume} {38}},\ \bibinfo {pages}
  {4265} (\bibinfo {year} {2013})}\BibitemShut {NoStop}%
\bibitem [{\citenamefont {Eppink}\ and\ \citenamefont
  {Parker}(1997)}]{Eppink_1997}%
  \BibitemOpen
  \bibfield  {author} {\bibinfo {author} {\bibfnamefont {A.~T. J.~B.}\
  \bibnamefont {Eppink}}\ and\ \bibinfo {author} {\bibfnamefont {D.~H.}\
  \bibnamefont {Parker}},\ }\href
  {https://doi.org/http://dx.doi.org/10.1063/1.1148310} {\bibfield  {journal}
  {\bibinfo  {journal} {Rev. Sci. Instrum.}\ }\textbf {\bibinfo {volume}
  {68}},\ \bibinfo {pages} {3477} (\bibinfo {year} {1997})}\BibitemShut
  {NoStop}%
\bibitem [{\citenamefont {Parker}\ and\ \citenamefont
  {Eppink}(1997)}]{Parker_1997}%
  \BibitemOpen
  \bibfield  {author} {\bibinfo {author} {\bibfnamefont {D.~H.}\ \bibnamefont
  {Parker}}\ and\ \bibinfo {author} {\bibfnamefont {A.~T. J.~B.}\ \bibnamefont
  {Eppink}},\ }\href {https://doi.org/http://dx.doi.org/10.1063/1.474624}
  {\bibfield  {journal} {\bibinfo  {journal} {J. Chem. Phys}\ }\textbf
  {\bibinfo {volume} {107}},\ \bibinfo {pages} {2357} (\bibinfo {year}
  {1997})}\BibitemShut {NoStop}%
\bibitem [{\citenamefont {Wollenhaupt}\ \emph {et~al.}(2009)\citenamefont
  {Wollenhaupt}, \citenamefont {Krug}, \citenamefont {K{\"o}hler},
  \citenamefont {Bayer}, \citenamefont {Sarpe-Tudoran},\ and\ \citenamefont
  {Baumert}}]{Wollenhaupt_2009}%
  \BibitemOpen
  \bibfield  {author} {\bibinfo {author} {\bibfnamefont {M.}~\bibnamefont
  {Wollenhaupt}}, \bibinfo {author} {\bibfnamefont {M.}~\bibnamefont {Krug}},
  \bibinfo {author} {\bibfnamefont {J.}~\bibnamefont {K{\"o}hler}}, \bibinfo
  {author} {\bibfnamefont {T.}~\bibnamefont {Bayer}}, \bibinfo {author}
  {\bibfnamefont {C.}~\bibnamefont {Sarpe-Tudoran}},\ and\ \bibinfo {author}
  {\bibfnamefont {T.}~\bibnamefont {Baumert}},\ }\href
  {https://doi.org/10.1007/s00340-009-3513-0} {\bibfield  {journal} {\bibinfo
  {journal} {Appl. Phys. B}\ }\textbf {\bibinfo {volume} {95}},\ \bibinfo
  {pages} {647} (\bibinfo {year} {2009})}\BibitemShut {NoStop}%
\bibitem [{\citenamefont {Smeenk}\ \emph {et~al.}(2009)\citenamefont {Smeenk},
  \citenamefont {Arissian}, \citenamefont {Staudte}, \citenamefont
  {Villeneuve},\ and\ \citenamefont {Corkum}}]{Smeenk_2009momentum}%
  \BibitemOpen
  \bibfield  {author} {\bibinfo {author} {\bibfnamefont {C.}~\bibnamefont
  {Smeenk}}, \bibinfo {author} {\bibfnamefont {L.}~\bibnamefont {Arissian}},
  \bibinfo {author} {\bibfnamefont {A.}~\bibnamefont {Staudte}}, \bibinfo
  {author} {\bibfnamefont {D.}~\bibnamefont {Villeneuve}},\ and\ \bibinfo
  {author} {\bibfnamefont {P.}~\bibnamefont {Corkum}},\ }\href@noop {}
  {\bibfield  {journal} {\bibinfo  {journal} {J. Phys. B}\ }\textbf {\bibinfo
  {volume} {42}},\ \bibinfo {pages} {185402} (\bibinfo {year}
  {2009})}\BibitemShut {NoStop}%
\bibitem [{\citenamefont {Dimitrovski}\ \emph {et~al.}(2014)\citenamefont
  {Dimitrovski}, \citenamefont {Maurer}, \citenamefont {Stapelfeldt},\ and\
  \citenamefont {Madsen}}]{Dimitrovski_2014}%
  \BibitemOpen
  \bibfield  {author} {\bibinfo {author} {\bibfnamefont {D.}~\bibnamefont
  {Dimitrovski}}, \bibinfo {author} {\bibfnamefont {J.}~\bibnamefont {Maurer}},
  \bibinfo {author} {\bibfnamefont {H.}~\bibnamefont {Stapelfeldt}},\ and\
  \bibinfo {author} {\bibfnamefont {L.~B.}\ \bibnamefont {Madsen}},\ }\href
  {https://doi.org/10.1103/PhysRevLett.113.103005} {\bibfield  {journal}
  {\bibinfo  {journal} {Phys. Rev. Lett.}\ }\textbf {\bibinfo {volume} {113}},\
  \bibinfo {pages} {103005} (\bibinfo {year} {2014})}\BibitemShut {NoStop}%
\bibitem [{\citenamefont {Nubbemeyer}\ \emph {et~al.}(2008)\citenamefont
  {Nubbemeyer}, \citenamefont {Gorling}, \citenamefont {Saenz}, \citenamefont
  {Eichmann},\ and\ \citenamefont {Sandner}}]{Nubbemeyer_2008}%
  \BibitemOpen
  \bibfield  {author} {\bibinfo {author} {\bibfnamefont {T.}~\bibnamefont
  {Nubbemeyer}}, \bibinfo {author} {\bibfnamefont {K.}~\bibnamefont {Gorling}},
  \bibinfo {author} {\bibfnamefont {A.}~\bibnamefont {Saenz}}, \bibinfo
  {author} {\bibfnamefont {U.}~\bibnamefont {Eichmann}},\ and\ \bibinfo
  {author} {\bibfnamefont {W.}~\bibnamefont {Sandner}},\ }\href@noop {}
  {\bibfield  {journal} {\bibinfo  {journal} {Phys. Rev. Lett.}\ }\textbf
  {\bibinfo {volume} {101}},\ \bibinfo {pages} {233001} (\bibinfo {year}
  {2008})}\BibitemShut {NoStop}%
\bibitem [{\citenamefont {Eichmann}\ \emph {et~al.}(2009)\citenamefont
  {Eichmann}, \citenamefont {Nubbemeyer}, \citenamefont {Rottke},\ and\
  \citenamefont {Sandner}}]{Eichmann_2009}%
  \BibitemOpen
  \bibfield  {author} {\bibinfo {author} {\bibfnamefont {U.}~\bibnamefont
  {Eichmann}}, \bibinfo {author} {\bibfnamefont {T.}~\bibnamefont
  {Nubbemeyer}}, \bibinfo {author} {\bibfnamefont {H.}~\bibnamefont {Rottke}},\
  and\ \bibinfo {author} {\bibfnamefont {W.}~\bibnamefont {Sandner}},\
  }\href@noop {} {\bibfield  {journal} {\bibinfo  {journal} {Nature}\ }\textbf
  {\bibinfo {volume} {461}},\ \bibinfo {pages} {1261} (\bibinfo {year}
  {2009})}\BibitemShut {NoStop}%
\bibitem [{\citenamefont {Perelomov}\ and\ \citenamefont {Popov}(1967)}]{PPT}%
  \BibitemOpen
  \bibfield  {author} {\bibinfo {author} {\bibfnamefont {A.~M.}\ \bibnamefont
  {Perelomov}}\ and\ \bibinfo {author} {\bibfnamefont {V.~S.}\ \bibnamefont
  {Popov}},\ }\href@noop {} {\bibfield  {journal} {\bibinfo  {journal} {Zh.
  Exp. Theor. Fiz.}\ }\textbf {\bibinfo {volume} {52}},\ \bibinfo {pages} {514}
  (\bibinfo {year} {1967})}\BibitemShut {NoStop}%
\bibitem [{\citenamefont {Ammosov}\ \emph {et~al.}(1986)\citenamefont
  {Ammosov}, \citenamefont {Delone},\ and\ \citenamefont {Krainov}}]{ADK}%
  \BibitemOpen
  \bibfield  {author} {\bibinfo {author} {\bibfnamefont {M.~V.}\ \bibnamefont
  {Ammosov}}, \bibinfo {author} {\bibfnamefont {N.~B.}\ \bibnamefont
  {Delone}},\ and\ \bibinfo {author} {\bibfnamefont {V.~P.}\ \bibnamefont
  {Krainov}},\ }\href@noop {} {\bibfield  {journal} {\bibinfo  {journal} {Zh.
  Eksp. Teor. Fiz.}\ }\textbf {\bibinfo {volume} {91}},\ \bibinfo {pages}
  {2008} (\bibinfo {year} {1986})}\BibitemShut {NoStop}%
\bibitem [{\citenamefont {Li}\ \emph {et~al.}(2014)\citenamefont {Li},
  \citenamefont {Geng}, \citenamefont {Liu}, \citenamefont {Deng},
  \citenamefont {Wu}, \citenamefont {Peng}, \citenamefont {Gong},\ and\
  \citenamefont {Liu}}]{Li_2014a}%
  \BibitemOpen
  \bibfield  {author} {\bibinfo {author} {\bibfnamefont {M.}~\bibnamefont
  {Li}}, \bibinfo {author} {\bibfnamefont {J.-W.}\ \bibnamefont {Geng}},
  \bibinfo {author} {\bibfnamefont {H.}~\bibnamefont {Liu}}, \bibinfo {author}
  {\bibfnamefont {Y.}~\bibnamefont {Deng}}, \bibinfo {author} {\bibfnamefont
  {C.}~\bibnamefont {Wu}}, \bibinfo {author} {\bibfnamefont {L.-Y.}\
  \bibnamefont {Peng}}, \bibinfo {author} {\bibfnamefont {Q.}~\bibnamefont
  {Gong}},\ and\ \bibinfo {author} {\bibfnamefont {Y.}~\bibnamefont {Liu}},\
  }\href@noop {} {\bibfield  {journal} {\bibinfo  {journal} {Phys. Rev. Lett.}\
  }\textbf {\bibinfo {volume} {112}},\ \bibinfo {pages} {113002} (\bibinfo
  {year} {2014})}\BibitemShut {NoStop}%
\bibitem [{\citenamefont {Shvetsov-Shilovski}\ \emph
  {et~al.}(2016)\citenamefont {Shvetsov-Shilovski}, \citenamefont {Lein},
  \citenamefont {Madsen}, \citenamefont {R\"as\"anen}, \citenamefont {Lemell},
  \citenamefont {Burgd\"orfer}, \citenamefont {Arb\'o},\ and\ \citenamefont
  {T\ifmmode~\mbox{\H{o}}\else \H{o}\fi{}k\'esi}}]{Shvetsov-Shilovski_2016}%
  \BibitemOpen
  \bibfield  {author} {\bibinfo {author} {\bibfnamefont {N.~I.}\ \bibnamefont
  {Shvetsov-Shilovski}}, \bibinfo {author} {\bibfnamefont {M.}~\bibnamefont
  {Lein}}, \bibinfo {author} {\bibfnamefont {L.~B.}\ \bibnamefont {Madsen}},
  \bibinfo {author} {\bibfnamefont {E.}~\bibnamefont {R\"as\"anen}}, \bibinfo
  {author} {\bibfnamefont {C.}~\bibnamefont {Lemell}}, \bibinfo {author}
  {\bibfnamefont {J.}~\bibnamefont {Burgd\"orfer}}, \bibinfo {author}
  {\bibfnamefont {D.~G.}\ \bibnamefont {Arb\'o}},\ and\ \bibinfo {author}
  {\bibfnamefont {K.}~\bibnamefont {T\ifmmode~\mbox{\H{o}}\else
  \H{o}\fi{}k\'esi}},\ }\href {https://doi.org/10.1103/PhysRevA.94.013415}
  {\bibfield  {journal} {\bibinfo  {journal} {Phys. Rev. A}\ }\textbf {\bibinfo
  {volume} {94}},\ \bibinfo {pages} {013415} (\bibinfo {year}
  {2016})}\BibitemShut {NoStop}%
\bibitem [{\citenamefont {Pfeiffer}\ \emph {et~al.}(2012)\citenamefont
  {Pfeiffer}, \citenamefont {Cirelli}, \citenamefont {Smolarski}, \citenamefont
  {Dimitrovski}, \citenamefont {Abu-samha}, \citenamefont {Madsen},\ and\
  \citenamefont {Keller}}]{Pfeiffer_2012}%
  \BibitemOpen
  \bibfield  {author} {\bibinfo {author} {\bibfnamefont {A.~N.}\ \bibnamefont
  {Pfeiffer}}, \bibinfo {author} {\bibfnamefont {C.}~\bibnamefont {Cirelli}},
  \bibinfo {author} {\bibfnamefont {M.}~\bibnamefont {Smolarski}}, \bibinfo
  {author} {\bibfnamefont {D.}~\bibnamefont {Dimitrovski}}, \bibinfo {author}
  {\bibfnamefont {M.}~\bibnamefont {Abu-samha}}, \bibinfo {author}
  {\bibfnamefont {L.~B.}\ \bibnamefont {Madsen}},\ and\ \bibinfo {author}
  {\bibfnamefont {U.}~\bibnamefont {Keller}},\ }\href@noop {} {\bibfield
  {journal} {\bibinfo  {journal} {Nature Phys.}\ }\textbf {\bibinfo {volume}
  {8}},\ \bibinfo {pages} {76} (\bibinfo {year} {2012})}\BibitemShut {NoStop}%
\bibitem [{\citenamefont {Popov}(2005)}]{Popov_2005}%
  \BibitemOpen
  \bibfield  {author} {\bibinfo {author} {\bibfnamefont {V.~S.}\ \bibnamefont
  {Popov}},\ }\href@noop {} {\bibfield  {journal} {\bibinfo  {journal} {Phys.
  Atom. Nuclei}\ }\textbf {\bibinfo {volume} {68}},\ \bibinfo {pages} {686}
  (\bibinfo {year} {2005})}\BibitemShut {NoStop}%
\bibitem [{\citenamefont {Landau}\ and\ \citenamefont
  {Lifshitz}(1975)}]{Landau_2}%
  \BibitemOpen
  \bibfield  {author} {\bibinfo {author} {\bibfnamefont {L.~D.}\ \bibnamefont
  {Landau}}\ and\ \bibinfo {author} {\bibfnamefont {E.~M.}\ \bibnamefont
  {Lifshitz}},\ }\href@noop {} {\emph {\bibinfo {title} {The Classical Theory
  of Fields}}}\ (\bibinfo  {publisher} {Elsevier, Oxford},\ \bibinfo {year}
  {1975})\BibitemShut {NoStop}%
\bibitem [{\citenamefont {Reiss}(1992)}]{Reiss_1992}%
  \BibitemOpen
  \bibfield  {author} {\bibinfo {author} {\bibfnamefont {H.}~\bibnamefont
  {Reiss}},\ }\href
  {https://doi.org/https://doi.org/10.1016/0079-6727(92)90008-J} {\bibfield
  {journal} {\bibinfo  {journal} {Prog. Quant. El.}\ }\textbf {\bibinfo
  {volume} {16}},\ \bibinfo {pages} {1 } (\bibinfo {year} {1992})}\BibitemShut
  {NoStop}%
\bibitem [{\citenamefont {Augst}\ \emph {et~al.}(1991)\citenamefont {Augst},
  \citenamefont {Meyerhofer}, \citenamefont {Strickland},\ and\ \citenamefont
  {Chin}}]{Augst_1991}%
  \BibitemOpen
  \bibfield  {author} {\bibinfo {author} {\bibfnamefont {S.}~\bibnamefont
  {Augst}}, \bibinfo {author} {\bibfnamefont {D.~D.}\ \bibnamefont
  {Meyerhofer}}, \bibinfo {author} {\bibfnamefont {D.}~\bibnamefont
  {Strickland}},\ and\ \bibinfo {author} {\bibfnamefont {S.~L.}\ \bibnamefont
  {Chin}},\ }\href@noop {} {\bibfield  {journal} {\bibinfo  {journal} {J. Opt.
  Soc. Am. B}\ }\textbf {\bibinfo {volume} {8}},\ \bibinfo {pages} {858}
  (\bibinfo {year} {1991})}\BibitemShut {NoStop}%
\bibitem [{\citenamefont {Brichta}\ \emph {et~al.}(2006)\citenamefont
  {Brichta}, \citenamefont {Liu}, \citenamefont {Zaidi}, \citenamefont
  {Trottier},\ and\ \citenamefont {Sanderson}}]{Brichta_2006}%
  \BibitemOpen
  \bibfield  {author} {\bibinfo {author} {\bibfnamefont {J.~P.}\ \bibnamefont
  {Brichta}}, \bibinfo {author} {\bibfnamefont {W.-K.}\ \bibnamefont {Liu}},
  \bibinfo {author} {\bibfnamefont {A.~A.}\ \bibnamefont {Zaidi}}, \bibinfo
  {author} {\bibfnamefont {A.}~\bibnamefont {Trottier}},\ and\ \bibinfo
  {author} {\bibfnamefont {J.~H.}\ \bibnamefont {Sanderson}},\ }\href@noop {}
  {\bibfield  {journal} {\bibinfo  {journal} {J. Phys. B}\ }\textbf {\bibinfo
  {volume} {39}},\ \bibinfo {pages} {3769} (\bibinfo {year}
  {2006})}\BibitemShut {NoStop}%
\bibitem [{\citenamefont {Rogers}(1981)}]{Rogers_1981}%
  \BibitemOpen
  \bibfield  {author} {\bibinfo {author} {\bibfnamefont {F.~J.}\ \bibnamefont
  {Rogers}},\ }\href@noop {} {\bibfield  {journal} {\bibinfo  {journal} {Phys.
  Rev. A}\ }\textbf {\bibinfo {volume} {23}},\ \bibinfo {pages} {1008}
  (\bibinfo {year} {1981})}\BibitemShut {NoStop}%
\bibitem [{\citenamefont {Milo{\v{s}}evi{\'{c}}}\ \emph
  {et~al.}(2009)\citenamefont {Milo{\v{s}}evi{\'{c}}}, \citenamefont {Becker},
  \citenamefont {Okunishi}, \citenamefont {Prümper}, \citenamefont {Shimada},\
  and\ \citenamefont {Ueda}}]{Milosevic_2009}%
  \BibitemOpen
  \bibfield  {author} {\bibinfo {author} {\bibfnamefont {D.~B.}\ \bibnamefont
  {Milo{\v{s}}evi{\'{c}}}}, \bibinfo {author} {\bibfnamefont {W.}~\bibnamefont
  {Becker}}, \bibinfo {author} {\bibfnamefont {M.}~\bibnamefont {Okunishi}},
  \bibinfo {author} {\bibfnamefont {G.}~\bibnamefont {Prümper}}, \bibinfo
  {author} {\bibfnamefont {K.}~\bibnamefont {Shimada}},\ and\ \bibinfo {author}
  {\bibfnamefont {K.}~\bibnamefont {Ueda}},\ }\href@noop {} {\bibfield
  {journal} {\bibinfo  {journal} {J. Phys. B}\ }\textbf {\bibinfo {volume}
  {43}},\ \bibinfo {pages} {015401} (\bibinfo {year} {2009})}\BibitemShut
  {NoStop}%
\bibitem [{\citenamefont {Liu}\ and\ \citenamefont
  {Hatsagortsyan}(2011)}]{Liu_2011}%
  \BibitemOpen
  \bibfield  {author} {\bibinfo {author} {\bibfnamefont {C.}~\bibnamefont
  {Liu}}\ and\ \bibinfo {author} {\bibfnamefont {K.~Z.}\ \bibnamefont
  {Hatsagortsyan}},\ }\href@noop {} {\bibfield  {journal} {\bibinfo  {journal}
  {J. Phys. B}\ }\textbf {\bibinfo {volume} {44}},\ \bibinfo {pages} {095402}
  (\bibinfo {year} {2011})}\BibitemShut {NoStop}%
\bibitem [{\citenamefont {K\"astner}\ \emph {et~al.}(2012)\citenamefont
  {K\"astner}, \citenamefont {Saalmann},\ and\ \citenamefont
  {Rost}}]{Kastner_2012}%
  \BibitemOpen
  \bibfield  {author} {\bibinfo {author} {\bibfnamefont {A.}~\bibnamefont
  {K\"astner}}, \bibinfo {author} {\bibfnamefont {U.}~\bibnamefont
  {Saalmann}},\ and\ \bibinfo {author} {\bibfnamefont {J.~M.}\ \bibnamefont
  {Rost}},\ }\href@noop {} {\bibfield  {journal} {\bibinfo  {journal} {Phys.
  Rev. Lett.}\ }\textbf {\bibinfo {volume} {108}},\ \bibinfo {pages} {033201}
  (\bibinfo {year} {2012})}\BibitemShut {NoStop}%
\bibitem [{\citenamefont {Arb{\'{o}}}\ \emph {et~al.}(2006)\citenamefont
  {Arb{\'{o}}}, \citenamefont {Persson},\ and\ \citenamefont
  {Burgd{\"{o}}rfer}}]{Arbo_2006a}%
  \BibitemOpen
  \bibfield  {author} {\bibinfo {author} {\bibfnamefont {D.~G.}\ \bibnamefont
  {Arb{\'{o}}}}, \bibinfo {author} {\bibfnamefont {E.}~\bibnamefont
  {Persson}},\ and\ \bibinfo {author} {\bibfnamefont {J.}~\bibnamefont
  {Burgd{\"{o}}rfer}},\ }\href@noop {} {\bibfield  {journal} {\bibinfo
  {journal} {Phys. Rev. A}\ }\textbf {\bibinfo {volume} {74}},\ \bibinfo
  {pages} {063407} (\bibinfo {year} {2006})}\BibitemShut {NoStop}%
\bibitem [{\citenamefont {Popov}(2004)}]{Popov_2004u}%
  \BibitemOpen
  \bibfield  {author} {\bibinfo {author} {\bibfnamefont {V.~S.}\ \bibnamefont
  {Popov}},\ }\href@noop {} {\bibfield  {journal} {\bibinfo  {journal} {Phys.
  Usp.}\ }\textbf {\bibinfo {volume} {47}},\ \bibinfo {pages} {855} (\bibinfo
  {year} {2004})}\BibitemShut {NoStop}%
\bibitem [{\citenamefont {Dan\ifmmode~\check{e}\else \v{e}\fi{}k}\ \emph
  {et~al.}(2018{\natexlab{c}})\citenamefont {Dan\ifmmode~\check{e}\else
  \v{e}\fi{}k}, \citenamefont {Hatsagortsyan},\ and\ \citenamefont
  {Keitel}}]{Danek_2018c}%
  \BibitemOpen
  \bibfield  {author} {\bibinfo {author} {\bibfnamefont {J.}~\bibnamefont
  {Dan\ifmmode~\check{e}\else \v{e}\fi{}k}}, \bibinfo {author} {\bibfnamefont
  {K.~Z.}\ \bibnamefont {Hatsagortsyan}},\ and\ \bibinfo {author}
  {\bibfnamefont {C.~H.}\ \bibnamefont {Keitel}},\ }\href@noop {} {\bibfield
  {journal} {\bibinfo  {journal} {Phys. Rev. A}\ }\textbf {\bibinfo {volume}
  {97}},\ \bibinfo {pages} {063410} (\bibinfo {year}
  {2018}{\natexlab{c}})}\BibitemShut {NoStop}%
\bibitem [{\citenamefont {Alnaser}\ \emph {et~al.}(2004)\citenamefont
  {Alnaser}, \citenamefont {Tong}, \citenamefont {Osipov}, \citenamefont
  {Voss}, \citenamefont {Maharjan}, \citenamefont {Shan}, \citenamefont
  {Chang},\ and\ \citenamefont {Cocke}}]{Alnaser_2004}%
  \BibitemOpen
  \bibfield  {author} {\bibinfo {author} {\bibfnamefont {A.~S.}\ \bibnamefont
  {Alnaser}}, \bibinfo {author} {\bibfnamefont {X.~M.}\ \bibnamefont {Tong}},
  \bibinfo {author} {\bibfnamefont {T.}~\bibnamefont {Osipov}}, \bibinfo
  {author} {\bibfnamefont {S.}~\bibnamefont {Voss}}, \bibinfo {author}
  {\bibfnamefont {C.~M.}\ \bibnamefont {Maharjan}}, \bibinfo {author}
  {\bibfnamefont {B.}~\bibnamefont {Shan}}, \bibinfo {author} {\bibfnamefont
  {Z.}~\bibnamefont {Chang}},\ and\ \bibinfo {author} {\bibfnamefont {C.~L.}\
  \bibnamefont {Cocke}},\ }\href {https://doi.org/10.1103/PhysRevA.70.023413}
  {\bibfield  {journal} {\bibinfo  {journal} {Phys. Rev. A}\ }\textbf {\bibinfo
  {volume} {70}},\ \bibinfo {pages} {023413} (\bibinfo {year}
  {2004})}\BibitemShut {NoStop}%
\bibitem [{\citenamefont {Smeenk}\ \emph
  {et~al.}(2011{\natexlab{b}})\citenamefont {Smeenk}, \citenamefont {Salvail},
  \citenamefont {Arissian}, \citenamefont {Corkum}, \citenamefont {Hebeisen},\
  and\ \citenamefont {Staudte}}]{Smeenk_2011oe}%
  \BibitemOpen
  \bibfield  {author} {\bibinfo {author} {\bibfnamefont {C.}~\bibnamefont
  {Smeenk}}, \bibinfo {author} {\bibfnamefont {J.~Z.}\ \bibnamefont {Salvail}},
  \bibinfo {author} {\bibfnamefont {L.}~\bibnamefont {Arissian}}, \bibinfo
  {author} {\bibfnamefont {P.~B.}\ \bibnamefont {Corkum}}, \bibinfo {author}
  {\bibfnamefont {C.~T.}\ \bibnamefont {Hebeisen}},\ and\ \bibinfo {author}
  {\bibfnamefont {A.}~\bibnamefont {Staudte}},\ }\href@noop {} {\bibfield
  {journal} {\bibinfo  {journal} {Optics Express}\ }\textbf {\bibinfo {volume}
  {19}},\ \bibinfo {pages} {9336} (\bibinfo {year}
  {2011}{\natexlab{b}})}\BibitemShut {NoStop}%
\bibitem [{\citenamefont {Friedrich}(2016)}]{Friedrich_Scattering}%
  \BibitemOpen
  \bibfield  {author} {\bibinfo {author} {\bibfnamefont {H.}~\bibnamefont
  {Friedrich}},\ }\href@noop {} {\emph {\bibinfo {title} {Scattering Theory}}}\
  (\bibinfo  {publisher} {Springer, Berlin, Heidelberg},\ \bibinfo {year}
  {2016})\BibitemShut {NoStop}%
\bibitem [{\citenamefont {Shvetsov-Shilovski}\ \emph
  {et~al.}(2009)\citenamefont {Shvetsov-Shilovski}, \citenamefont
  {Goreslavski}, \citenamefont {Popruzhenko},\ and\ \citenamefont
  {Becker}}]{Shvetsov-Shilovski_2009}%
  \BibitemOpen
  \bibfield  {author} {\bibinfo {author} {\bibfnamefont {N.}~\bibnamefont
  {Shvetsov-Shilovski}}, \bibinfo {author} {\bibfnamefont {S.}~\bibnamefont
  {Goreslavski}}, \bibinfo {author} {\bibfnamefont {S.}~\bibnamefont
  {Popruzhenko}},\ and\ \bibinfo {author} {\bibfnamefont {W.}~\bibnamefont
  {Becker}},\ }\href@noop {} {\bibfield  {journal} {\bibinfo  {journal} {Laser
  Physics}\ }\textbf {\bibinfo {volume} {19}},\ \bibinfo {pages} {1550}
  (\bibinfo {year} {2009})}\BibitemShut {NoStop}%
\end{thebibliography}%

\end{document}